\documentclass{aastex}
\usepackage{amsmath}
\usepackage{amsfonts}
\usepackage{amssymb}
\usepackage{amsthm}
\usepackage[english]{babel}
\usepackage{textcomp}
\usepackage{natbib}
\usepackage{graphicx}
\usepackage{lineno}
\usepackage{lscape}
\usepackage{color}
\linenumbers

\newcommand{\g}{$\gamma$}
\newcommand{\hi}{$\mathrm{H\,\scriptstyle{I}}$}

\newcommand{\kms}{km s$^{-1}$}
\newcommand{\msol}{\hbox{$M_\odot$}}                   
\newcommand{\sdeg}{$^{\circ}$}                         
\newcommand{\ddeg}{\hbox{$.\!\!^\circ$}}               
\newcommand{\TS}{{$TS$ }}

\makeatletter

\newcommand{\Rmnum}[1]{\expandafter\@slowromancap\romannumeral #1@}
\makeatother


\begin{document}

\title{$Fermi$ LAT and $WMAP$ observations of the supernova remnant HB~21}
\author{
G.~Pivato\altaffilmark{1,2,3},
J.~Hewitt\altaffilmark{4,5,6}, 
L.~Tibaldo\altaffilmark{7,8} 
F.~Acero\altaffilmark{5}, 
J.~Ballet\altaffilmark{9}, 
T.~J.~Brandt\altaffilmark{5}, 
F.~de~Palma\altaffilmark{10,11}, 
F.~Giordano\altaffilmark{10,11}, 
G.H~Janssen\altaffilmark{12},  
G.~J\'ohannesson\altaffilmark{13}, 
D.~A.~Smith\altaffilmark{14}
}
\altaffiltext{1}{Dipartimento di Fisica e Astronomia ``G. Galilei'', Universit\`a di Padova, I-35131 Padova, Italy}
\altaffiltext{2}{Istituto Nazionale di Fisica Nucleare, Sezione di Padova,
I-35131 Padova, Italy}
\altaffiltext{3}{email: giovanna.pivato@pd.infn.it}
\altaffiltext{4}{CRESST, University of Maryland, Baltimore County, Baltimore, MD 21250}
\altaffiltext{5}{NASA Goddard Space Flight Center, Greenbelt, MD 20771, USA}
\altaffiltext{6}{email: john.w.hewitt@nasa.gov}
\altaffiltext{7}{W. W. Hansen Experimental Physics Laboratory, Kavli Institute for Particle Astrophysics and Cosmology, Department of Physics and SLAC National Accelerator Laboratory, Stanford University, Stanford, CA 94305, USA}
\altaffiltext{8}{email: ltibaldo@slac.stanford.edu}
\altaffiltext{9}{Laboratoire AIM, CEA-IRFU/CNRS/Universit\'e Paris Diderot, Service d'Astrophysique, CEA Saclay, 91191 Gif sur Yvette, France}
\altaffiltext{10}{Dipartimento di Fisica ``M. Merlin" dell'Universit\`a e del Politecnico di Bari, I-70126 Bari, Italy}
\altaffiltext{11}{Istituto Nazionale di Fisica Nucleare, Sezione di Bari, 70126 Bari, Italy}
\altaffiltext{12}{University of Manchester, Manchester, M13 9PL, UK}
\altaffiltext{13}{Science Institute, University of Iceland, IS-107 Reykjavik, Iceland}
\altaffiltext{14}{Centre d'\'Etudes Nucl\'eaires de Bordeaux Gradignan, IN2P3/CNRS, Universit\'e Bordeaux 1, BP120, F-33175 Gradignan Cedex, France}

\begin{abstract}
We present the analysis of \textit{Fermi}~Large Area Telescope (LAT) \g-ray
observations
of HB~21 (G89.0+4.7). We detect
significant \g-ray emission associated with the remnant: the flux
$>100$~MeV is
$9.4\pm0.8\;(\mathrm{stat})\pm1.6\;(\mathrm{syst})\times10^{-11}$ erg
cm$^{-2}$ s$^{-1}$. HB~21 is well
modeled by a uniform disk
centered at $l$ = 88\ddeg75 $\pm$ 0\ddeg04, $b$ = +4\ddeg65 $\pm$ 0\ddeg06 with a radius
of 1\ddeg19 $\pm$ 0\ddeg06.
The
\g-ray spectrum shows clear evidence of curvature, suggesting a cutoff or
break in the underlying particle population at an energy of a few GeV.
We complement \g-ray observations with the analysis of the {\it WMAP} 7-year data from
23 to 93 GHz, achieving the first detection of HB~21 at these frequencies. 
In combination with archival radio data, the radio spectrum shows a spectral break
which helps to constrain the relativistic electron spectrum, hence parameters
of simple non-thermal radiation models. In one-zone models multiwavelength data
favor the origin of \g~rays from nucleon-nucleon collisions. A
single population
of electrons cannot produce both \g~rays through
bremsstrahlung and radio emission through synchrotron radiation.
A predominantly inverse-Compton origin of the \g-ray emission is disfavored
because it requires lower interstellar densities than are inferred for HB~21. 
In the hadronic-dominated scenarios accelerated nuclei 
contribute a total energy of $\sim 3 \times10^{49}$ ergs, while in
a two-zone bremsstrahlung-dominated scenario the total energy in
accelerated
particles is $\sim$1$\times10^{49}$ ergs.\end{abstract}

\section{Introduction}\label{intro}

Diffusive shock acceleration in supernova remnants (SNRs) is strongly advocated
as the mechanism responsible for the acceleration of  Galactic cosmic rays (CRs,
e.g.~\citealt{drury12}). High-energy emission from SNRs is a powerful probe of
this process \citep{rey08}. In particular, high-energy \g-ray emission can
pinpoint the presence of energetic leptons or ions, constraining the
acceleration efficiency and the maximum energy of accelerated particles
\citep[e.g.][]{rxhess,rxferm,tyver,tyferm}.

SNRs interacting with dense molecular clouds are expected to be
bright \g-ray sources \citep[e.g.,][]{hew09}.
Archetypal remnants of this class are IC~443 and W44,
for which GeV- and, for the former, TeV-energy \g-ray emission are observed and
interpreted as
the signature of collisions between accelerated nuclei and interstellar matter
\citep{ic443ferm,ic443mag,w44ferm}, as recently confirmed by the
detection of the characteristic spectral feature of \g-rays produced from
the decay of neutral pions \citep{w44ag,fermipidec}.

Both sources belong to a class of objects known as mixed-morphology
(MM) SNRs, where an outer shell revealed by non-thermal radio emission surrounds
a central region filled by thermal gas emitting in X-rays \citep{rho98}. There
is a strong association between MM SNRs and evidence for
interactions between shock fronts and dense clouds revealed by
OH masers \citep{zadeh03}. Therefore, MM  SNRs are good candidates to be
bright \g-ray sources. Indeed, in addition to W44 and IC~443, other members of
this class are known to be GeV \g-ray emitters: W28 \citep{w28ferm}, W49B
\citep{w49bfermi}, W51C \citep{w51cfermi} and 3C~391 \citep{3c391fermi}.

HB~21 (also known as G89.0+4.7, \citealt{green}) is a MM SNR interacting with
molecular clouds. Evidence for interaction comes from several different wavelengths.
X-ray observations with \emph{Einstein} provide evidence that
the SN explosion took place in a low-density cavity, therefore suggesting a
massive stellar progenitor \citep{kno96}. HB~21 has a radio shape with bright filaments and
indentations suggesting interactions with the nearby clouds
\citep{Byu06}. Looking at the CO distribution
it is possible to distinguish two different structures based on the velocity of
the molecular clouds. \citet{koo91} report the presence of broad CO lines
emitted from shocked clumps of molecular gas with high column densities.
Infrared line observations
\citep{shi09} reveal an infrared spectrum from H$_2$ gas with a mixture of 
temperatures, indicating shocks into a multi-phase medium, typical of molecular clouds.

The distance to HB~21 was first estimated to be $0.8$~kpc on the basis of
an association with the Cygnus OB7 complex
and ROSAT observations of X-ray absorption \citep{lea96}. Considering the
preshock velocities, the X-ray absorbing column density, the \hi\ and CO
distribution along the line of sight and the
highly polarized emission, the distance was later determined to be
$\sim$1.7 kpc \citep{Byu06}. In this work we adopt the later, revised distance.

The age of the SNR is also uncertain. \citet{laz06} determine an age of
$\sim$5600 years by fitting evolutionary models to the thermal X-ray spectrum
of the SNR. However, this results from assuming that
the SNR is located at 0.8 kpc, and still in the adiabatic phase.
Kinematics of the \hi\ shell velocity of 124~km~s$^{-1}$
indicates a much older age for the remnant of $4.5\times10^4$~yr \citep{koo91}.
Additionally, H$_2$ spectra indicate steady-state continuous shocks, which
require $\sim$10$^{4}$ yr to develop \citep{flo99}. In at least some regions
optical filaments are observed with line ratios typical of velocities $\sim$150
\kms, indicating that the shock has evolved into the
radiative phase \citep{mavromatakis07}.
A SNR shock propagating into a uniform medium of $\sim$1 cm$^{-3}$ becomes
radiative at an age of $\sim 3.6 \times10^{4}$ yr \citep{mavromatakis07}.
Given this evidence, we
argue that HB~21 is a 
middle-aged SNR, at an age of a few tens of thousands of years, similar to other
SNRs interacting with molecular clouds detected by \emph{Fermi} Large Area Telescope (LAT).

High-energy \g-ray sources coincident with HB~21 were reported
in the 2-year Catalog of the LAT on board
the \emph{Fermi Gamma-ray Space Telescope} \citep{2fgl}.
Three sources are associated with the remnant (2FGL~J2043.3+5105, J2046.0+4954,
and J2041.5+5003) and an additional source is present on the edge
(2FGL~J2051.8+5054): we will investigate if it is part of the emission of the
remnant or it is a source itself.
\citet{reichardt12} recently reported an analysis of 3.5 years of
\textit{Fermi}~LAT data towards
HB~21, finding evidence for extended emission and claiming a softer spectrum in
the direction of a shocked cloud in the north-western region of the
remnant.

Due to the large apparent size ($\sim$2\sdeg) this object is well suited for a
detailed morphological study using \emph{Fermi}~LAT data.
Therefore, it is ideal to further assess the correlation between
cloud-interacting shocks and \g-ray emission as a tracer of particle
acceleration in MM SNRs, as well as to investigate the possible spectral
variations that could shed light on the poorly known processes of particle
escape from the shocks \citep[e.g.][]{gabici09}.

In this paper we present the
analysis of $\sim4$~years of LAT observations of HB~21 and we discuss the \g-ray
emission mechanism in light of a new analysis of {\it WMAP}
7-years data, as well as of archival radio data. In \S\ref{gammas} we
describe the \g-ray observations and the morphological and spectral
characterization of
the \g-ray emission from HB~21. In \S\ref{sec:radio} we present the analysis of the 
{\it WMAP} and archival radio data, which reveals a break in the high-frequency
radio spectrum.
In \S\ref{discussion} the interpretation of these results is discussed.

\section{Gamma-ray analysis}\label{gammas}

\subsection{Observations}
\label{observations}

The \emph{Fermi}~LAT is a pair-conversion \g-ray telescope detecting photons
from
20~MeV to $>300$~GeV \citep{Atw09}. Its on-orbit calibration is described in
\citet{abd09} and \cite{bal12} along with the event classification and
instrument performance. 

For this analysis, data are accumulated from the beginning of scientific
operations on 2008 August 4 to 2012 June 14, selecting the low-background 
P7SOURCE event class. The data analysis was performed using the LAT \emph{Science Tools} package (\texttt{v9r27p1}), available from the \emph{Fermi} Science Support Center\footnote{\url{http://fermi.gsfc.nasa.gov/ssc}}. For the morphological characterization we use
only events with energy $>$1~GeV
to profit from the narrower point-spread function (PSF) in order to separate the
\g-ray emission associated with HB~21 from neighboring sources and interstellar
emission. We then use events down to $100$~MeV to determine the spectral energy
distribution of the remnant.
Below this energy the PSF becomes much broader than the SNR and the
uncertainties related to the instrument response are larger. In both the
morphological and spectral characterization we consider photons with measured
energies up to $300$~GeV, but only find a significant detection of the source up
to energies of several GeV due to the limited number of events at high
energies.

We perform the analysis in a 10\sdeg$\times$10\sdeg\ region of interest
(RoI) centered at the radio position of HB~21 ($l$ = 89\ddeg0, $b$ = +4\ddeg7).
The RoI approximately corresponds to the 68\% event containment
region for the P7SOURCE events at 100~MeV and exceeds the 95\%
event containment region for energies $\gtrsim 700$~MeV. We
adopted this narrower than usual RoI in order to limit at the lowest energies
the large uncertainties due to the modeling of the bright interstellar
emission from the nearby Cygnus region \citep{Cyg_cocoon,LATCygISM2012}. We also
exclude a few time intervals
when the LAT boresight was
rocked with respect to the local zenith by more than 52\sdeg\ (mostly for
calibration purposes or to point at specific sources) and events with a
reconstructed angle with respect to the local zenith $>$100\sdeg\ in order to
limit the contribution from the Earth's atmospheric \g-ray emission. To take
into account the latter selection criterion in the calculation of the exposure,
we exclude time intervals when any part of the RoI was observed at zenith angles
$>$100\sdeg .

\subsection{Background model and analysis method}
\label{background}

The background is composed of diffuse emission and individual nearby \g-ray
sources. Diffuse emission is taken into account using the standard models
provided by the \emph{Fermi} LAT
collaboration\footnote{\url{
http://fermi.gsfc.nasa.gov/ssc/data/access/lat/BackgroundModels.html}} for the
P7SOURCE selection \citep[see][]{2fgl}. They include a model that accounts for
the Galactic interstellar emission from CR interactions with interstellar gas
and low-energy radiation fields, and an isotropic background spectrum that
accounts for diffuse \g-ray emission of extragalactic origin and residual
background events due to charged particle interactions in the LAT misclassified
as \g-rays. We leave the normalization of the Galactic interstellar model as a
free parameter, yet fix the isotropic background spectrum because it is
difficult to separate it from the other components in such a small RoI and it is
more reliably determined over larger regions of the sky.

We include in the background model all the point sources present in the 2FGL
catalog \citep{2fgl} with distances less than 15\sdeg\ from the RoI center and
which are not associated with the SNR. We will discuss in \S\ref{morphology} the
case of the source 2FGL~J2051.8+5054, which is located at the edge of the SNR.
The spectral models used for background sources are those reported in the 2FGL
catalog. Fluxes and spectral indices are left as free parameters in the fit if
the source is within the RoI. Otherwise they are fixed to the catalog values.

The background model and the various models for HB~21 are fitted to the LAT
data using a binned maximum-likelihood method based on Poisson statistics
\citep[e.g.][]{Mat96}, as implemented in the \emph{gtlike} tool. For this
purpose data were binned on an angular grid with 0\ddeg1 spacing, and different
binning in energy as detailed below. The analysis uses the post-launch
instrument response functions (IRFs) P7SOURCE$\_$V6 \citep{bal12}.

Finally, we note that the radio pulsar J2047+5029 is 0\ddeg5 away from the remnant's radio center \citep{pul09}. 
While pulsars represent the largest Galactic \g-ray source class, this particular pulsar likely contributes no detectable flux to the ROI. 
The spindown power is $\dot{E}\sim2\times10^{33}$ erg s$^{-1}$, lower than that of any known \g-ray pulsar,
and three times lower than for any young, radio-loud \g-ray pulsar
\citep{2PC}. 
That is, the pulsar is probably below the ``deathline'' expected from the outer magnetospheric emission models that best
describe the LAT pulsars \citep{wan11}.
Furthermore, the pulsar's dispersion measure indicates a distance $D = 4.4$~kpc.
The ratio $\sqrt{\dot{E}}/D^2 = 2.3\times10^{15}$ erg$^{1/2}$ kpc$^{-2}$
(a proxy of the expected \g-ray luminosity) is
five times lower than for any known \g-ray pulsar \citep[see Fig.
15 in ][]{2PC}.
The two-year LAT catalog shows no point source at the pulsar position \citep{2fgl},
nor does the 4-year catalog currently in preparation.
Nevertheless, we searched PSR J2047+5029 for \g-ray pulsations. 
It is not part of \textit{Fermi}'s pulsar timing campaign \citep{smi08}, 
so we obtained a timing solution based on Westerbork Synthesis Radio Telescope (SRT) and Jodrell Bank radio data taken concurrently with the \emph{Fermi} data\footnote{Janssen, and Stappers private communication}.
The ephemeris will be presented in future work (Janssen et al. 2013). 
We used it to phase-fold the LAT data, over a grid of minimum energy cuts
(100~MeV to 1000 MeV, in 50 MeV steps)
and maximum radius cuts (from $0\fdg4$ to $2\arcdeg$ from the
pulsar position, in $0\fdg1$ steps).
For pulsars with unknown \g-ray spectral shapes and unknown pulse profile shapes, such grids amount to a search
for the maximum pulsar signal-to-background noise ratio. At each grid location, we calculated the H-test, which never
exceeded $3 \sigma$ statistical significance \citep[see Section 5 in ][]{2PC}.
The pulsar's energy flux above 100 MeV is therefore lower than for any known \g-ray pulsar, that is, below $2\times10^{-12}$ erg cm$^{-2}$ s$^{-1}$.
We conclude that the pulsar can be neglected when characterizing \g-ray emission
from HB~21.

\subsection{Systematic uncertainties}
\label{sys}

Two major sources of systematic errors on the results are the uncertainties in
the LAT effective area and the modeling of interstellar emission.
The uncertainties in the effective area for the IRFs we use are
evaluated as $10\%$ at $100$~MeV, $5\%$ at $516$~MeV, and $10\%$ above $10$~GeV,
linearly varying with the logarithm of energy between those values
\citep{bal12}. We estimate the error induced in the characterization of the
\g-ray fluxes of HB~21 by repeating the analysis with two sets
of
modified IRFs where the effective area was upscaled or downscaled by its
uncertainty.
This approach neglects errors induced by possibly more complicated
variations of the effective area and, therefore, does not capture uncertainties
in the spectral shape of the source. However, this is sufficient for our
purpose since, as discussed in section \ref{spectral}, for HB~21 systematic
uncertainties are dominated by the modeling of interstellar emission.

To gauge the systematic uncertainties due to the interstellar emission model we
compare the results obtained using the standard model in \S\ref{background} with
the results based on eight alternative interstellar emission models. The
alternative emission models are based on a subsample of those examined by
\citet{diffpapII}. We varied some of the most important parameters of the
interstellar emission models, namely the uniform spin temperature used to
estimate the column densities of interstellar atomic hydrogen ($150$~K and
$10^5$~K, the latter being equivalent to the small optical depth approximation),
the vertical height of the CR propagation halo ($4$~kpc and $10$~kpc), the CR
source distribution in the Galaxy (the pulsar distribution by
\citealt{lorimer2006} and the SNR distribution by
\citealt{case1998}\footnote{The validity of the
\citet{case1998} $\sigma-D$ relation has been criticised: it is used
in \citet{diffpapII} as an alternative to probe the effect of changing the CR
distribution.}). In this
way we built a grid of eight alternative interstellar emission models. We refer
the reader to \citet{diffpapII} for further details about the modeling strategy
and multimessenger/multiwavelength constraints on these parameters.

Each of these models was fitted to the whole-sky P7SOURCE data in the energy
range between $100$~MeV and $300$~GeV using the maximum likelihood method
described in \citet{diffpapII}. We determine independent log-parabolic
re-normalization functions separately for the Inverse Compton (IC) component and the
components associated to atomic hydrogen and molecular hydrogen (traced by CO),
split along the line of sight into four broad Galactocentric annuli between
0--4~kpc, 4--8~kpc, 8--10~kpc (the local annulus) and 10--30~kpc. 

The eight resulting models were used to repeat the source analysis described in
\S\ref{morphology}, for the determination of the disk dimension, and
\S\ref{spectral} for the spectral analysis of the source. In each of the
iterations we replaced the standard isotropic background and Galactic
interstellar emission models with the alternative isotropic background spectra
and Galactic interstellar emission templates, independently for the
sub-components described above (note that for the line of sight of HB~21 only
the two outermost Galactocentric annuli are relevant). For the same reasons as
for the standard model, we fix the isotropic background spectrum. We leave free
the
normalization parameters of the interstellar model components in the third and
fourth ring, and of the IC component. Through this procedure we
explored the impact of adopting a different model-construction strategy,
of varying the model parameters and of allowing more 
freedom in the fit. We stress that we explored only some of the uncertainties
related to the modeling of interstellar emission, but we are nevertheless able
to obtain an indication about some  important systematic effects. A more
thorough discussion on this method for exploring the systematic uncertainties
due to the modeling of diffuse emission is available in \citet{depalma13}.

\subsection{Morphological analysis}
\label{morphology}

\begin{figure}[htbp]
 \centering
\begin{tabular}{lr}
 \includegraphics[width=0.5\textwidth]{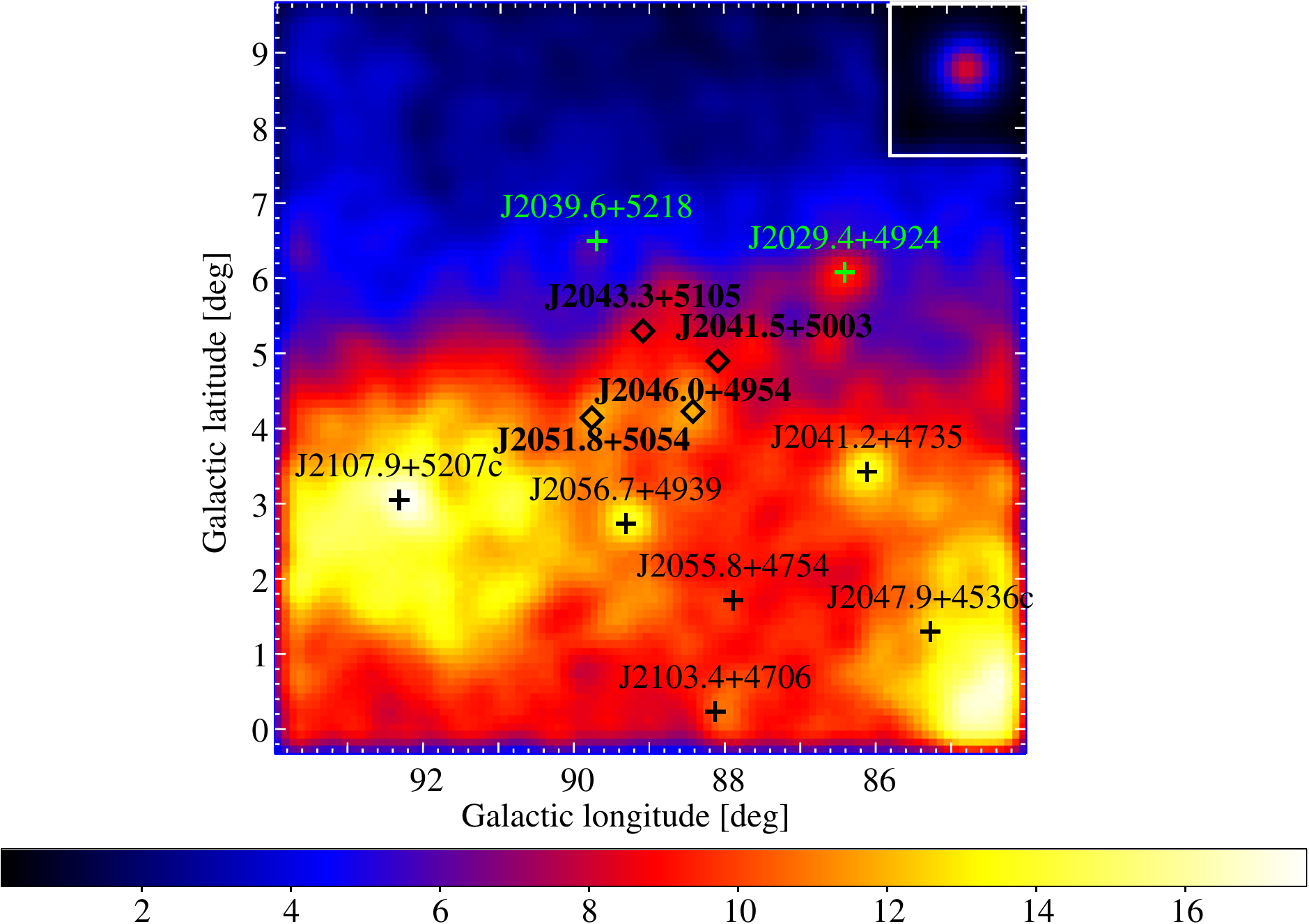} 
\put(-180,150){\hbox{\small{\texttt{\color{white}{\textbf{(a)}}}}}}
\put(-65,140){\hbox{\tiny{\texttt{\color{white}{\textbf{PSF}}}}}} &
\includegraphics[width=0.5\textwidth]{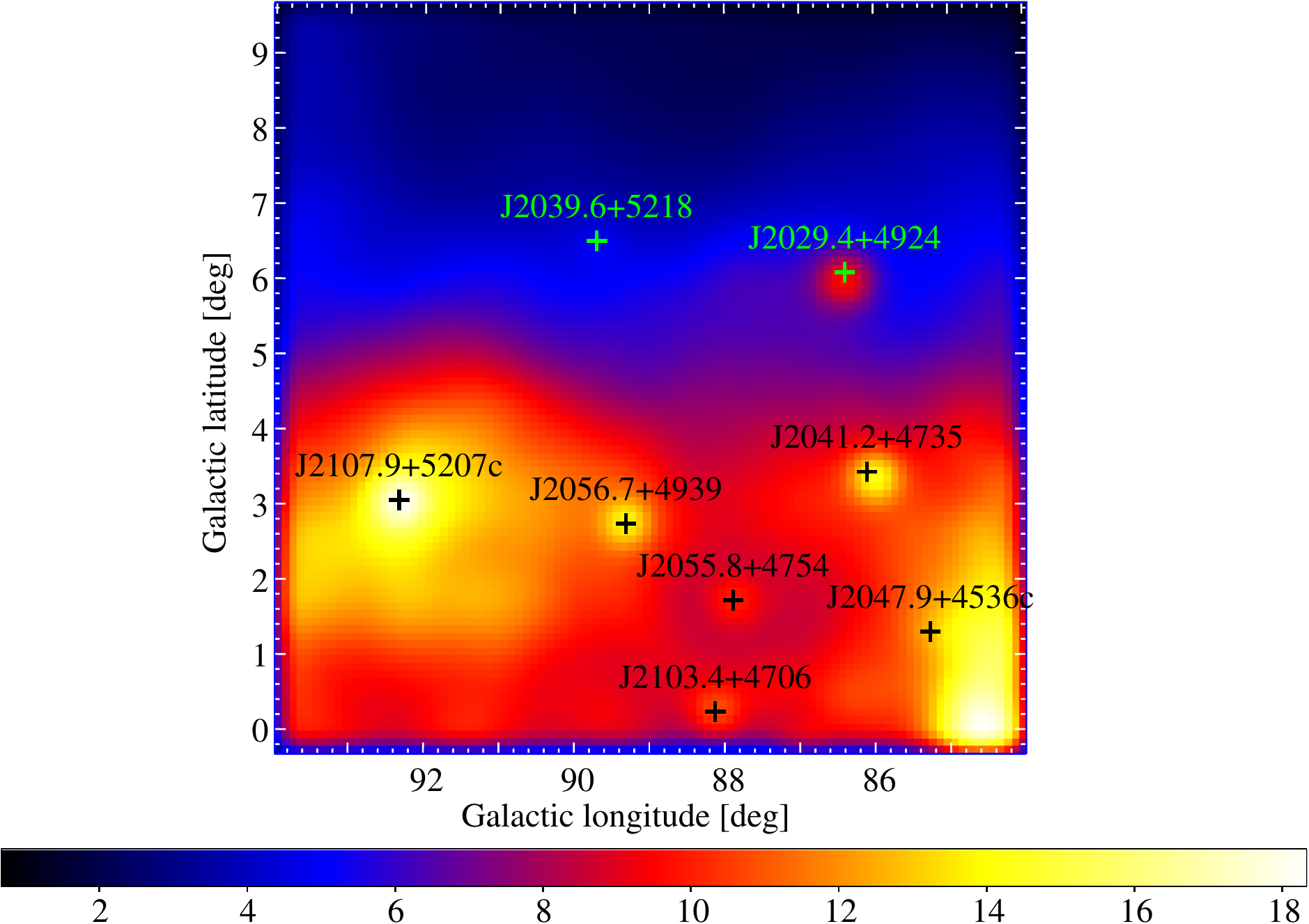} 
\put(-180,150){\hbox{\small{\texttt{\color{white}{\textbf{(b)}}}}}}\\
\includegraphics[width=0.5\textwidth]{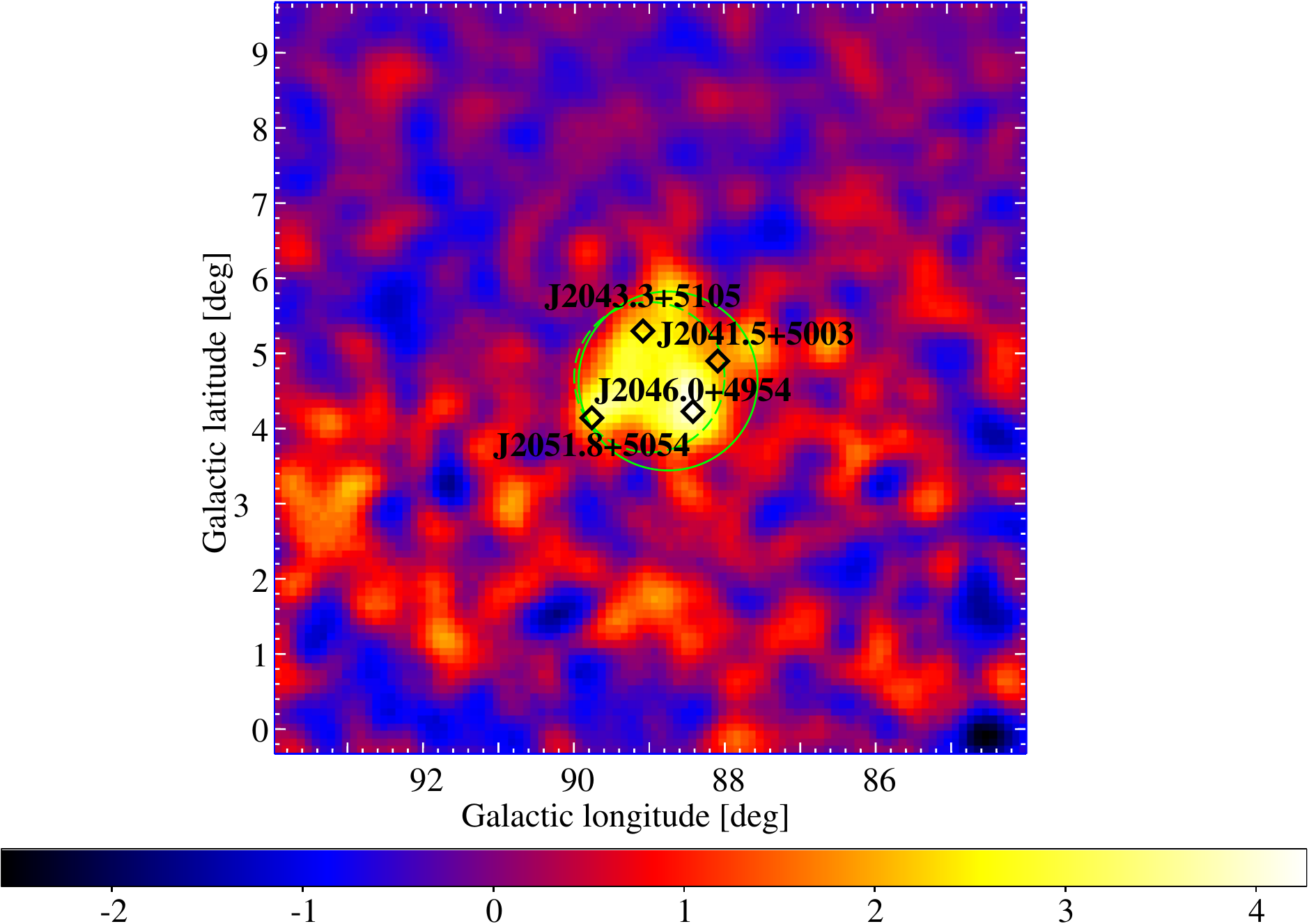}
\put(-180,150){\hbox{\small{\texttt{\color{white}{\textbf{(c)}}}}}} & \\
\end{tabular}
 \caption{a) LAT counts map at energies $> 1$~GeV with color scale
in counts/pixel. We overlaid the positions
of 2FGL sources (crosses for background
 sources and bold diamonds for the three sources associated with the remnant and
2FGL~J2051.8+5054). The inset in the top right corner shows the effective PSF
over the energy range considered for a power-law spectral distribution with
index 3.1. 
b) Background model map (calculated using the best-fit parameters
from the case where HB~21 was modeled as a disk).  
c) Remaining emission associated with HB~21; overlaid are the positions of  the
four point sources above, the best-fit disk (solid line) and the radio disk (dashed line).
The pixel size is 0\ddeg1 and all maps are smoothed for display with a Gaussian
kernel of
 $\sigma$=0\ddeg4.}
 \label{maps}
 \end{figure}

In Figure~\ref{maps}a we show a count map of the RoI for energies $>$1~GeV,
to visually illustrate the morphology of the \g-ray emission
in the region. We consider different spatial models for the emission from HB~21,
summarized in Table~\ref{spatial_ll}. For each model we evaluate the test
statistic
\begin{equation}
 TS=2(\ln \mathcal{L}_1-\ln \mathcal{L}_0)
\end{equation}
where $\mathcal{L}_1$ is the maximum-likelihood value for the model including
the remnant
and $\mathcal{L}_0$ is the maximum-likelihood value for the model not including
it (null
hypothesis). If the null hypothesis is verified (no \g-ray emission associated
to HB~21), \TS is expected to be distributed as a $\chi^2$ with a number of
degrees of freedom given by the additional number of free parameters in the
model including the source, with the caveats discussed in \citet{Protassov02}.

We determine the position and extension of the \g-ray emission associated with
HB~21 from \g-ray data only using the \emph{pointlike} tool, which is optimized
and widely validated against Monte Carlo simulations for this purpose
\citep{lan12}. We model the source as a disk with a power-law spectrum and we
determine the position of its center $(l,b)$ and radius $r$, along with flux and
spectral index from LAT data, using an energy binning of eight bins per decade.
We considered both the cases where 2FGL~J2051.8+5054 is
included as a separate point source in the model or removed. In the
first case we obtain the best fit parameters:
$l$ = 88\ddeg62 $\pm$ 0\ddeg05, $b$ = 4\ddeg79 $\pm$ 0\ddeg06, and
$r$ = 1\ddeg14 $\pm$ 0\ddeg07. In the second we have $l$ = 88\ddeg75 $\pm$ 0\ddeg04,
$b$ = 4\ddeg65 $\pm$ 0\ddeg06, $r$ = 1\ddeg19 $\pm$ 0\ddeg06. Errors reflect
statistical uncertainties only. The \TS for the case including separately
2FGL~J2051.8+5054 in the model is larger by $14$ than the case with the disk
only (Table~\ref{spatial_ll}). The significance of the separate source
hypothesis is below the threshold usually required to claim a detection for LAT
sources (\TS~$>25$). Deeper observations are needed to determine if
2FGL~J2051.8+5054 can be distinguished as a source independent from the
remnant. In the rest of the paper, we do not consider 2FGL~J2051.8+5054 as a separate
source and, therefore, we remove it from the model.
We note that the results concerning the morphology of HB~21 are
robust regardless of whether
2FGL~J2051.8+5054 is included as a separate source in the model or not.
Using the best fit disk, the significance of the detection of HB~21
is $\sim29~\sigma$. We also
check whether the source size changes with energy by separately fitting
a disk to the LAT data from 1 to 3~GeV and from 3 to 10~GeV. We obtain
$r$ = 1\ddeg19 $\pm$ 0\ddeg09 in the lower-energy range and
$r$ = 1\ddeg24 $\pm$ 0\ddeg09 in the higher-energy range. No significant change in
size with energy is detected.

In Figure~\ref{maps}b we show the background model map obtained from the fit
with the disk, and in Figure~\ref{maps}c the background-subtracted count map
where the emission associated to HB~21 is visible. 
If we compare the \TS with the model in which HB~21 is modeled by the four point
sources in the 2FGL Catalog (Table~\ref{spatial_ll}) we find that the extended
source provides a higher likelihood for a lower number of free parameters
in the fit. Therefore, we conclude that the hypothesis of extended emission
is preferred over four individual point sources.

We then calculate the systematic uncertainties related to interstellar emission using the alternative models as described in \S \ref{sys}.
When the alternative interstellar emission models are used, the \g-ray disk is
systematically shifted toward the north-west part with respect to the radio
shell (with shifts in longitude between 0\ddeg19 and
0\ddeg24, and in latitude between 0\ddeg06 and 0\ddeg09), and the disk radius is
systematically
smaller by 0\ddeg18--0\ddeg24, but the significance of the
detection of HB~21 does not change sizably. This effect is mainly due
to
the different approaches used to deal
with the dark gas, neutral interstellar gas which is not well traced by the linear combination of
column densities inferred from the 21-cm \hi\ line and the 2.6-mm CO line, but which is traced by
correlated residuals in dust emission/absorption and interstellar \g-ray emission
\citep{grenier2005}. Dust residuals are used as a dark-gas template fitted to the \g-ray data in
the standard interstellar model, whereas they are used to correct the \hi\ column densities in
the alternative models assuming a dust-to-gas ratio independent from \g-ray observations. This leads to
different estimates of the gas column densities, therefore
to different structures in the interstellar emission models. Additionally, the
standard model accounts
for enhanced interstellar emission toward the nearby Cygnus~X complex
\citep[see][]{Cyg_cocoon,LATCygISM2012},
whereas the alternative models do not. These differences are found to have a significant impact on
the determination of the SNR morphology. We note that for all the alternative
models the
disk extends beyond the rim of the remnant in coincidence with the western
molecular cloud, but leave the faint south-east edge of the radio shell off.

\begin{deluxetable}{ccc}
\tablecaption{Test statistic (\TS) and degrees of freedom (d.o.f.) for
the different spatial models for the \g-ray emission associated to
HB~21 considered in~\ref{morphology}.}
\tablenum{1}
\tablewidth{0pt}
\tablehead{\colhead{sources} & \colhead{\TS} & \colhead{d.o.f.} \\ 
\colhead{} & \colhead{} & \colhead{} } 
\startdata
Null hypothesis & 0 & 0 \\
4 point sources & 256 & 10 \\
disk & 302 & 5\\
disk + 2FGL~J2051.8+5054 & 316 & 7\\
X-ray image & 212 & 2 \\
X-ray image+ 2FGL~J2051.8+5054 & 234 & 4\\
radio image & 298 & 2 \\
radio image + 2FGL~J2051.8+5053 & 312 & 4 \\
\enddata
\label{spatial_ll}
\end{deluxetable}

\begin{figure}[htbp]
\centering
\begin{tabular}{lr}
\includegraphics[width=0.39\textwidth]{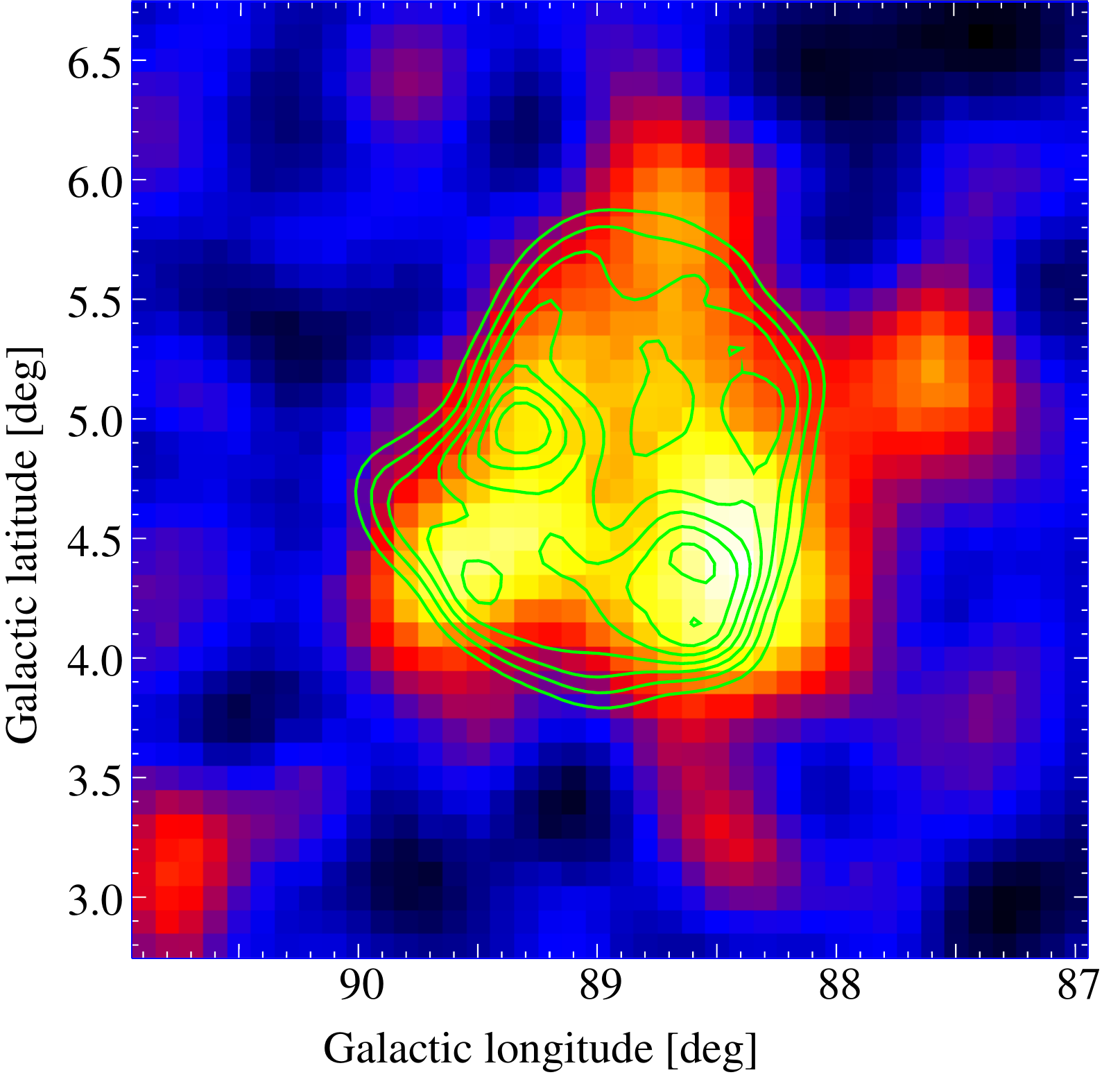}  
\put(-160,160){\hbox{\large{\texttt{\color{white}{\textbf{(a)}}}}}} &
\includegraphics[width=0.39\textwidth]{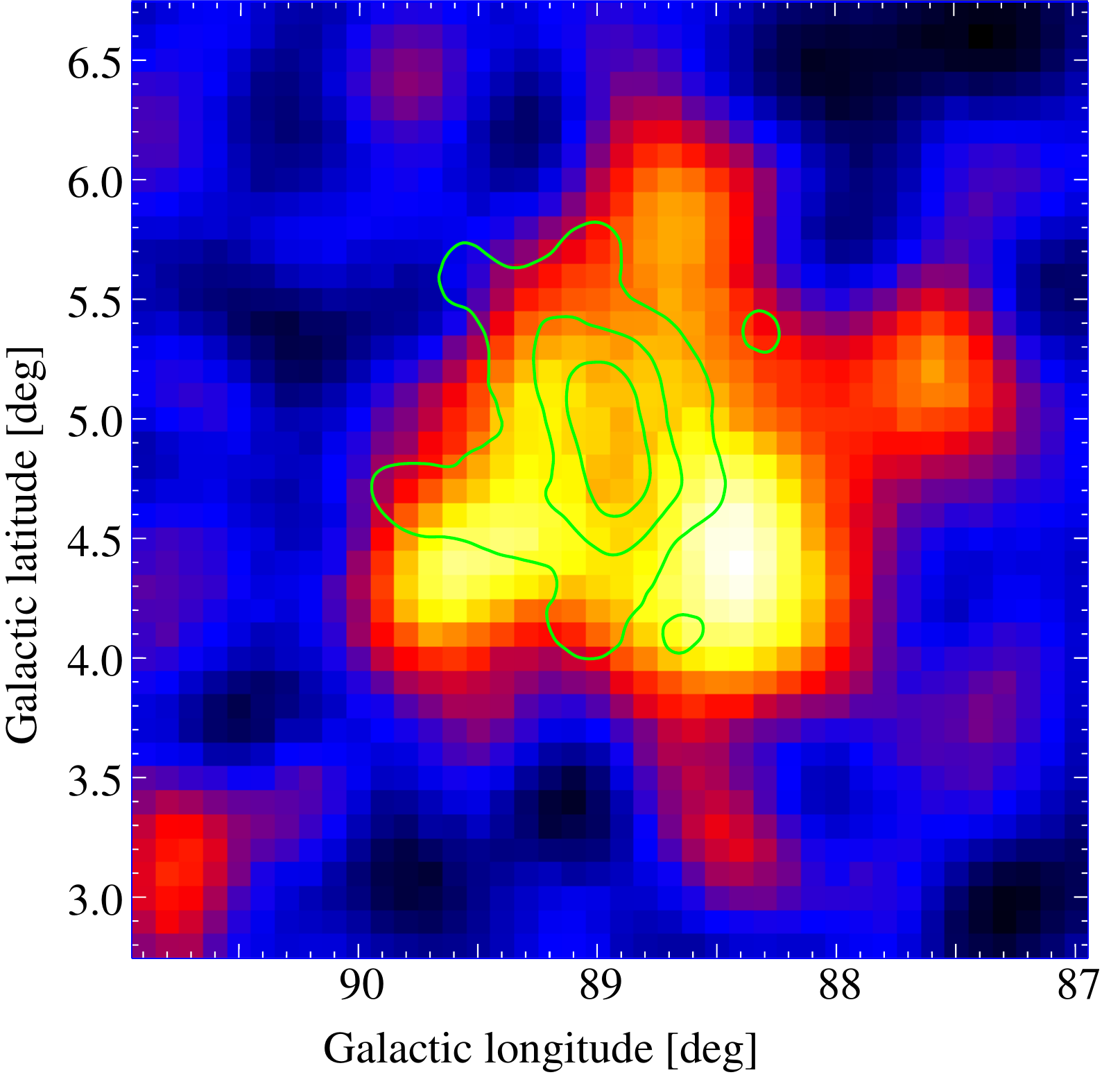} 
\put(-160,160){\hbox{\large{\texttt{\color{white}{\textbf{(b)}}}}}} \\
\includegraphics[width=0.39\textwidth]{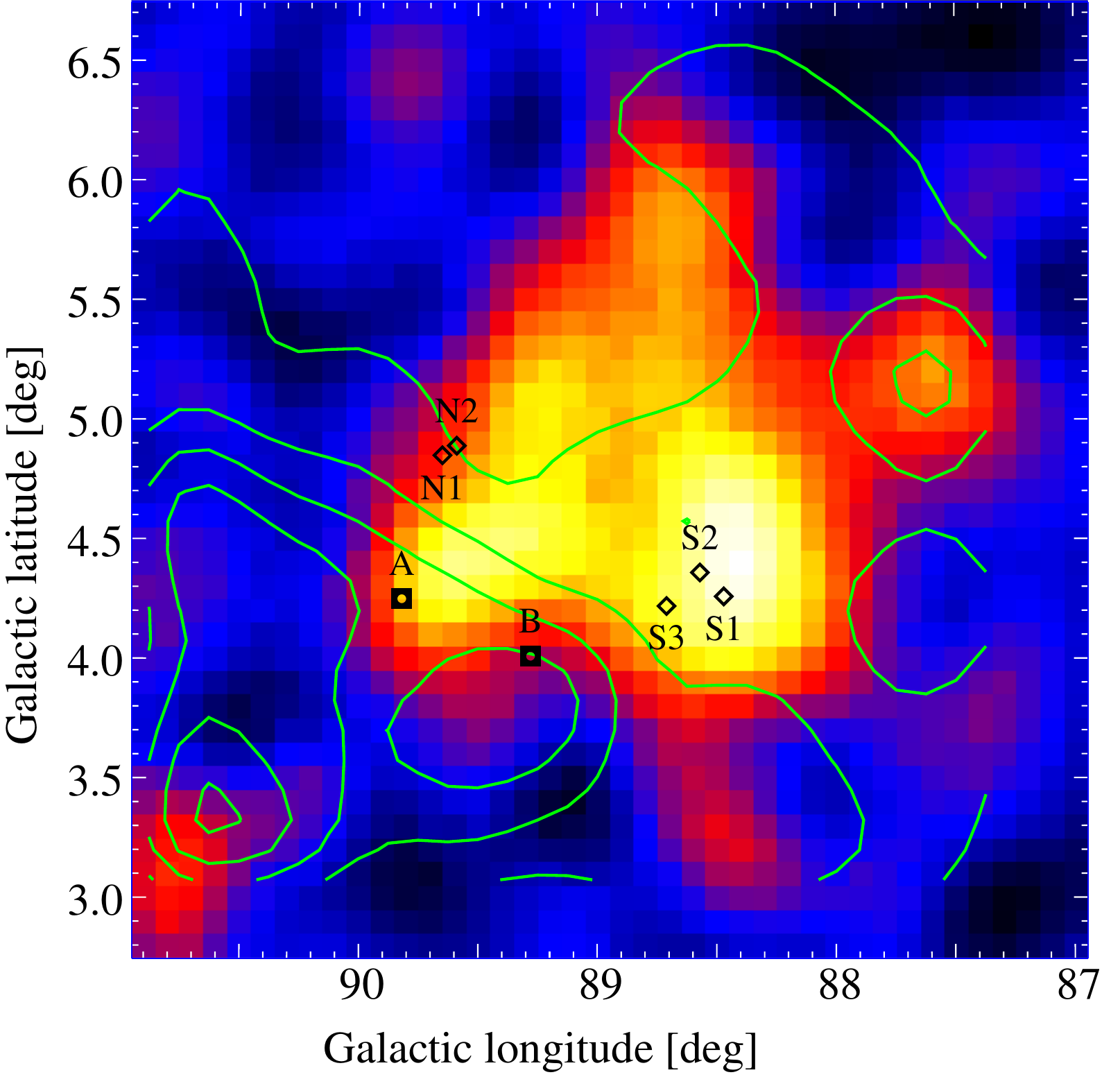} 
\put(-160,160){\hbox{\large{\texttt{\color{white}{\textbf{(c)}}}}}} & \\
\includegraphics[width=0.39\textwidth]{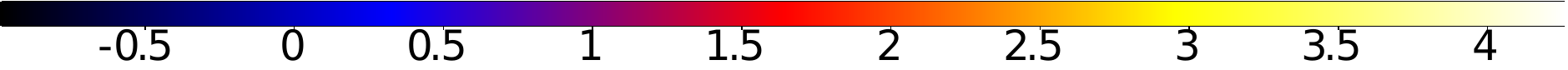} & \\
\end{tabular}
\caption{Emission associated with HB~21 (Fig.~\ref{maps}c) overlaid with
contours from emission at other wavelengths (see text for details):
a) Radio emission intensity at 6-cm with a beam size of 0\ddeg16,  smoothed with a Gaussian kernel with $\sigma$=0\ddeg2.
The seven contour levels are linearly spaced from 0 to 2.0 Jy beam$^{-1}$. 
b) Background-subtracted X-ray emission (ROSAT) smoothed with a Gaussian
kernel of $\sigma$ = 0\ddeg25. The three contour levels are linearly
spaced from 0.36$\times10^{-3}$ to 2.13$\times10^{-3}$
counts~s$^{-1}$~arcmin$^{-2}$.
c) Intensities of the 2.6-mm CO line in the Local-Arm region. The six contour
levels are linearly spaced from 1.5~K~km~s$^{-1}$ to 28~K~km~s$^{-1}$. We also
show the positions of the shocked CO clumps (diamonds) and clouds A and B
(squares) given in
\citet{koo91}. }
\label{over} 
\end{figure}

In Figure~\ref{over} we compare the \g-ray emission associated with HB~21 to emission from the remnant at other wavelengths. 
Figure~\ref{over}a compares the \g-ray image with radio emission at 6-cm tracing
non-thermal electrons (see later \S\ref{sec:radio}). Figure~\ref{over}b compares
the \g-ray morphology with X-ray emission from the thermal plasma filling the
center of the remnant, measured by ROSAT \citep{rosat}. The ROSAT image was
cleaned using the standard background maps\footnote{Available from
\url{http://www.xray.mpe.mpg.de/cgi-bin/rosat/rosat-survey}}.  Finally
Figure~\ref{over}c compares the \g-ray emitting region with the distribution of
molecular material.  Molecular column densities in the Local-Arm region are
traced by the CO line intensities at 2.6-mm \citep{dame01, dame11} integrated
over velocities with respect to the local standard of rest within $\pm
20$~km~s$^{-1}$. We also show the positions of 
shocked CO clumps and of two large CO clouds distributed along the boundary of the radio shell \citep{koo91}. 
The \g-ray emission associated with HB~21 is broader than the central region filled by thermal X-ray emitting plasma. It compares well with the radio shell, but appears to extend beyond the radio shell in regions where molecular clouds are present. The brightest \g-ray regions coincide with the Southern shocked CO clumps identified by \citet{koo91}. This correlation with CO suggests that at least part of the \g-ray emission from HB~21 may be produced by accelerated particles colliding with dense interstellar matter.

To quantitatively assess the correlation between \g-ray, X-ray and radio emission from HB~21 we use the X-ray and radio images as templates in the \g-ray analysis. For both templates we compare the cases when 2FGL~J2051.8+5054 is included in the fit as a background source or not.
All the spatial templates are fitted with a power-law spectrum. From the
results reported in Table~\ref{spatial_ll} we confirm that the morphology of the
remnant in \g-rays significantly differs, as expected, from the emission in X-rays. 
Both the radio template and the \g-ray disk provide a good fit to
the data, though the first has less degrees of freedom, as shown in Table
\ref{spatial_ll}. Since the two models are not nested it is not possible to
conclude which is the best one. Given that the disk model provides the largest \TS of all single source models, and is derived from the \g-ray data, we use this model to compute the spectrum of the remnant.

\subsection{Spectral analysis}
\label{spectral}

The spectral energy distribution (SED) of HB~21 is determined over the full
energy range between 100~MeV and 300~GeV (using 35
logarithmic energy bins for the likelihood analysis) by
modeling the remnant with the best-fit disk. We compare the hypotheses of a
power law
\begin{equation}
   \frac{dN}{dE}=N_0\left(\frac{E}{100 \mbox{ MeV}}\right)^{-\Gamma}
\end{equation}
with three different curved spectral models in order to quantify deviations from
the former. The three functional forms are:
\begin{enumerate}
	\item log-parabola.
		\begin{equation}
 			 \frac{dN}{dE}=N_0\left(\frac{E}{1000\mbox{ MeV}}\right)^{-\left(\alpha+\beta\ln(E/1000\mbox{\footnotesize{ MeV}})\right)}
 		 \label{logp}
		\end{equation}
	\item smooth broken power law
		\begin{equation}
   			\frac{dN}{dE}=N_0\left(\frac{E}{100}\right)^{-\Gamma_1}\left(1+\left(\frac{E}{E_b}\right)^{(\Gamma_2-\Gamma_1)/0.2}\right)^{-0.2}
		\label{sbpwl}
		\end{equation}
with $\Gamma_1$ and $\Gamma_2$ being the spectral indices below and above the
break energy $E_b$, respectively
	\item power law with exponential cutoff
		\begin{equation}
   			\frac{dN}{dE}=N_0\left(\frac{E}{200\mbox{ MeV}}\right)^{-\Gamma}\exp\left(-\frac{E}{E_c}\right)
		\label{plsc}
		\end{equation}
where $E_c$ is the cutoff energy.
\end{enumerate}

\begin{deluxetable}{cccc}
\tablecaption{\TS\ and additional degrees of freedom for the different
functions used to model the SED of HB~21.  For each model we report the
values of the best-fit parameters.}
\tablenum{2}
\tablewidth{0pt}
\tablehead{\colhead{spectrum shape} & \colhead{$\Delta$\TS} &
\colhead{additional d.o.f.} & \colhead{Fit parameters} \\
\colhead{} & \colhead{} & \colhead{}  & \colhead{}  }  
\startdata
 power law & 0 & 0 & \g=2.33$\pm$0.03 \\
 log-parabola & 92 & 1 &  $\alpha=2.54\pm0.05$ \\
 & & & $\beta=0.39\pm0.04$ \\
 smooth broken power law & 41 & 2 & \g$_1$=1.67$\pm$0.02\\
 & & & \g$_2$=3.54$\pm$0.05 \\
 & & & E$_b$=789$\pm$65 MeV  \\
 power law with exponential cutoff & 82 & 1 & \g$_1$=1.42$\pm$0.03 \\
 & & & E$_{c}$=958$\pm$41 MeV  \\
 \enddata
\label{spec_type}
\end{deluxetable}

In Table \ref{spec_type} we report the \TS values and the additional degrees of
freedom for the curved spectral models compared to the power law. All the
curved models have a higher \TS than the power law distribution. For the
following discussion we adopt the log-parabola because it yields the largest
improvement with respect to the simple power-law, with a  $\Delta TS=92$,
corresponding to an improvement at a $\sim$9$\sigma$ significance level in the energy range from 100~MeV to 300~GeV.
The total \g-ray energy flux from HB~21 results to be
$9.4\pm0.8\;(\mathrm{stat})\pm1.6\;(\mathrm{syst})\times10^{-11}$ erg
cm$^{-2}$ s$^{-1}$ and the photon flux  $1.48\pm0.2\;(\mathrm{stat})\pm0.4\;(\mathrm{syst})\times10^{-7}$ ph cm$^{-2}$ s$^{-1}$.
The systematic errors shown in this section are calculated as described in~\S\ref{sys} and extracting the root mean square of the variations with respect to the values
from the standard model. 

We also computed the SED in a model-independent way by splitting the energy
range between 0.1~MeV and 60~GeV in 12 logarithmically spaced bins. The model is the same as described above, but
in each narrow energy bin we leave free only the overall fluxes; other
spectral parameters are fixed to the 2FGL values for background sources
and to a power-law index of~2 for HB~21 (the results are insensitive to this
particular choice). When \TS for an individual bin is $<9$
we calculate an upper limit at the 95\% confidence level determined through the
likelihood profile method.

\begin{figure}[htbp]
 \centering
\includegraphics[width=\textwidth]{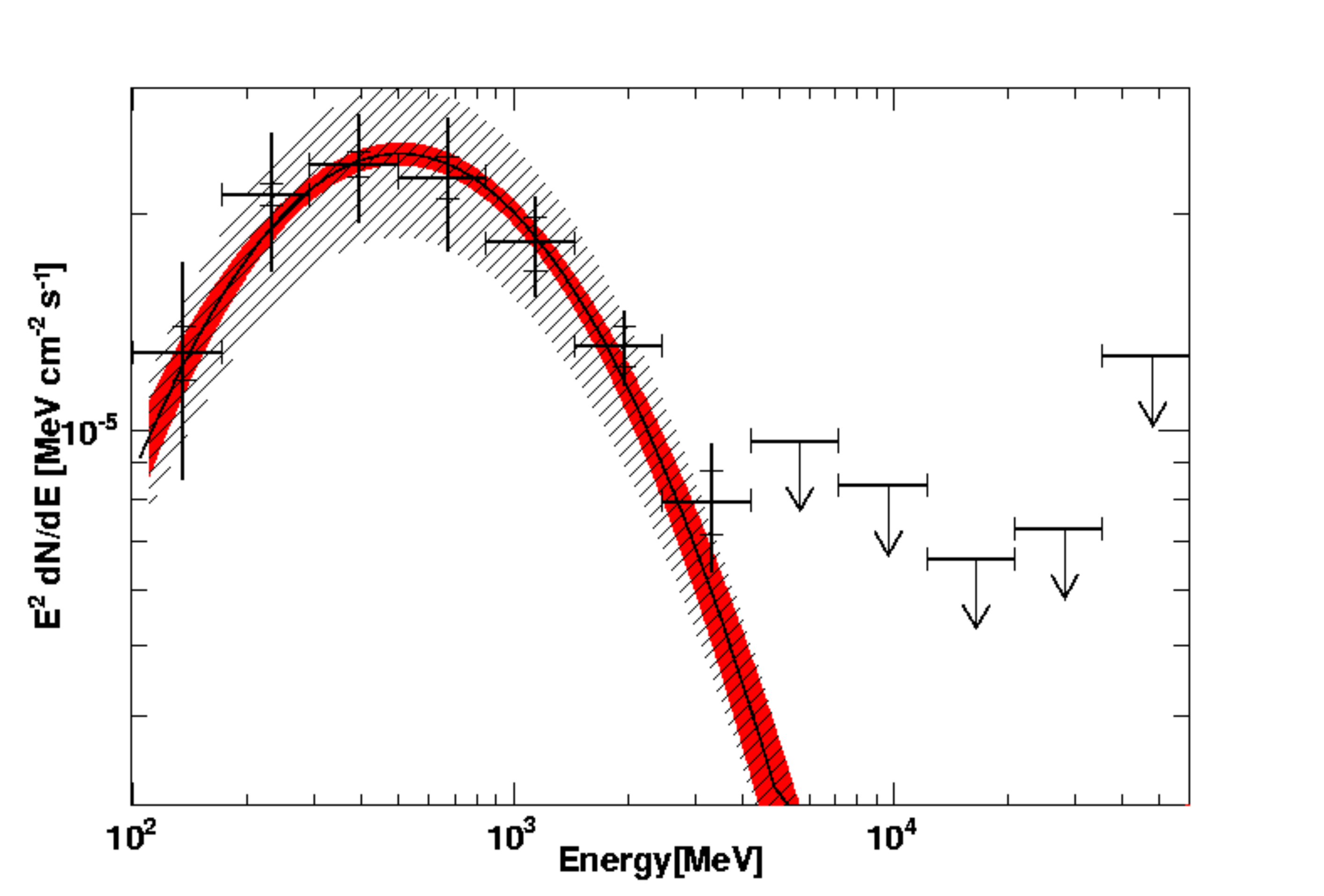}
 \caption{Spectral energy distribution of HB~21. The
line shows the best-fit log-parabola
 model, the light-red filled area shows the statistical error band  and the
dashed gray area shows
 the systematic uncertainties. The bar markers correspond to statistical errors only, while lines show the larger systematic errors. 95\%
confidence-level upper limits are given for energy bins where the \TS of the
source is $<9$.}
 \label{sed}
 \end{figure}

We show the resulting SED in Figure~\ref{sed} with also systematic errors indicated. Systematic errors due to interstellar emission model are then summed in quadrature with the
error due to the LAT effective area uncertainties (\S\ref{sys}) for display.
The SED points are reported in Table \ref{sysSED}. The contributions due to
the modeling of interstellar emission and the LAT effective area are presented
separately. In the energy range considered, the systematic errors are driven by the interstellar emission model.

\begin{deluxetable}{cccccc}
\tablecaption{SED of \g-ray emission from HB~21.
For each SED point we report the systematic uncertainties separately for those
related to interstellar emission modeling (IEM) and LAT effective area (EA),
as well as the overall uncertainty obtained by summing them in quadrature.}
\tablenum{3}
\tablewidth{0pt}
\tablehead{\colhead{\small{Energy bin center}} & \colhead{$E^2\mathrm{d}N/\mathrm{d}E$} & \colhead{\small{Statistical}} & \colhead{\small{IEM}} &  \colhead{\small{EA}} &\colhead{\small{Total syst.}} \\
\colhead{\footnotesize{[MeV]}}&\colhead{\footnotesize{[eV~cm$^{-2}$~s$^{-1}$]}}&\colhead{\footnotesize{[eV~cm$^{-2}$~s$^{-1}$]}} &\colhead{\footnotesize{[eV~cm$^{-2}$~s$^{-1}$]} } &\colhead{\footnotesize{[eV~cm$^{-2}$~s$^{-1}$]}} &\colhead{\footnotesize{[eV~cm$^{-2}$~s$^{-1}$]}} }  
\startdata
135 & 12.8 & $\pm$1.1 & $\pm$4.1 & $\pm$1.2 & $\pm$4.3 \\
230 & 21.3 & $\pm$0.8 & $\pm$4.3 & $\pm$1.8 & $\pm$4.7 \\
393 & 23.5 & $\pm$0.9 & $\pm$3.8 & $\pm$1.6 & $\pm$4.1 \\
670 & 22.5 & $\pm$1.5 & $\pm$4.6 & $\pm$1.1 & $\pm$4.8 \\
1143 & 18.3 & $\pm$1.5 & $\pm$2.7 & $\pm$0.9 & $\pm$2.9 \\
1950 & 13.1 & $\pm$0.9 & $\pm$1.4 & $\pm$0.7 & $\pm$1.6 \\
3327 & 7.9 & $\pm$0.8 & $\pm$1.5 & $\pm$0.5 & $\pm$1.6 \\
 \enddata
\label{sysSED}
\end{deluxetable}

\subsection{Search for spectral variations across the remnant}\label{specvar}

\begin{figure}[htbp]
 \centering
 \begin{tabular}{lr}
 \includegraphics[width=0.3\textwidth]{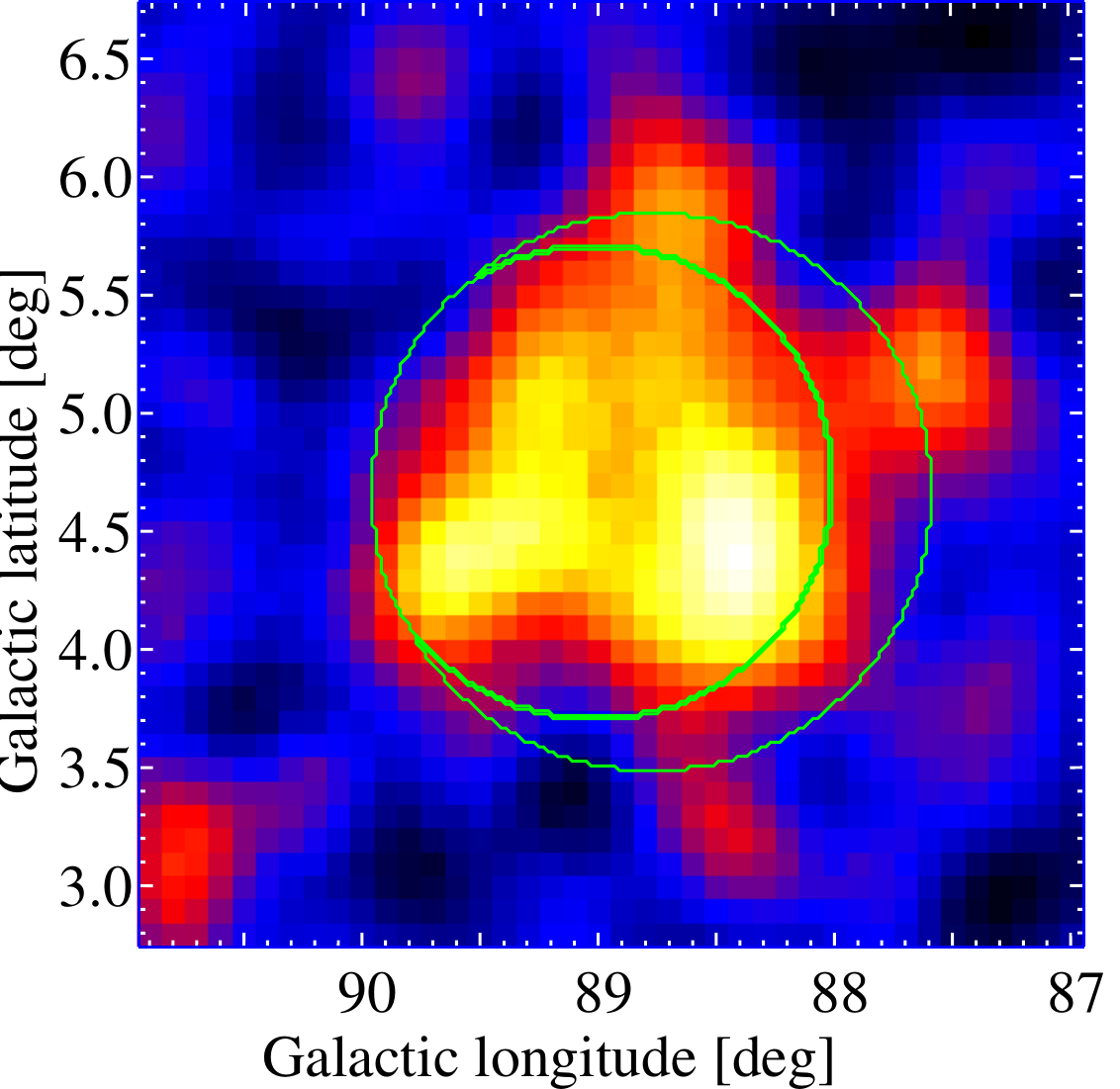}
\put(-120,120){\hbox{\large{\texttt{\color{white}{\textbf{(a)}}}}}}
\put(-80,75){\hbox{\footnotesize{\texttt{\color{black}{\textbf{radio}}}}}}
&
  \includegraphics[width=0.3\textwidth]{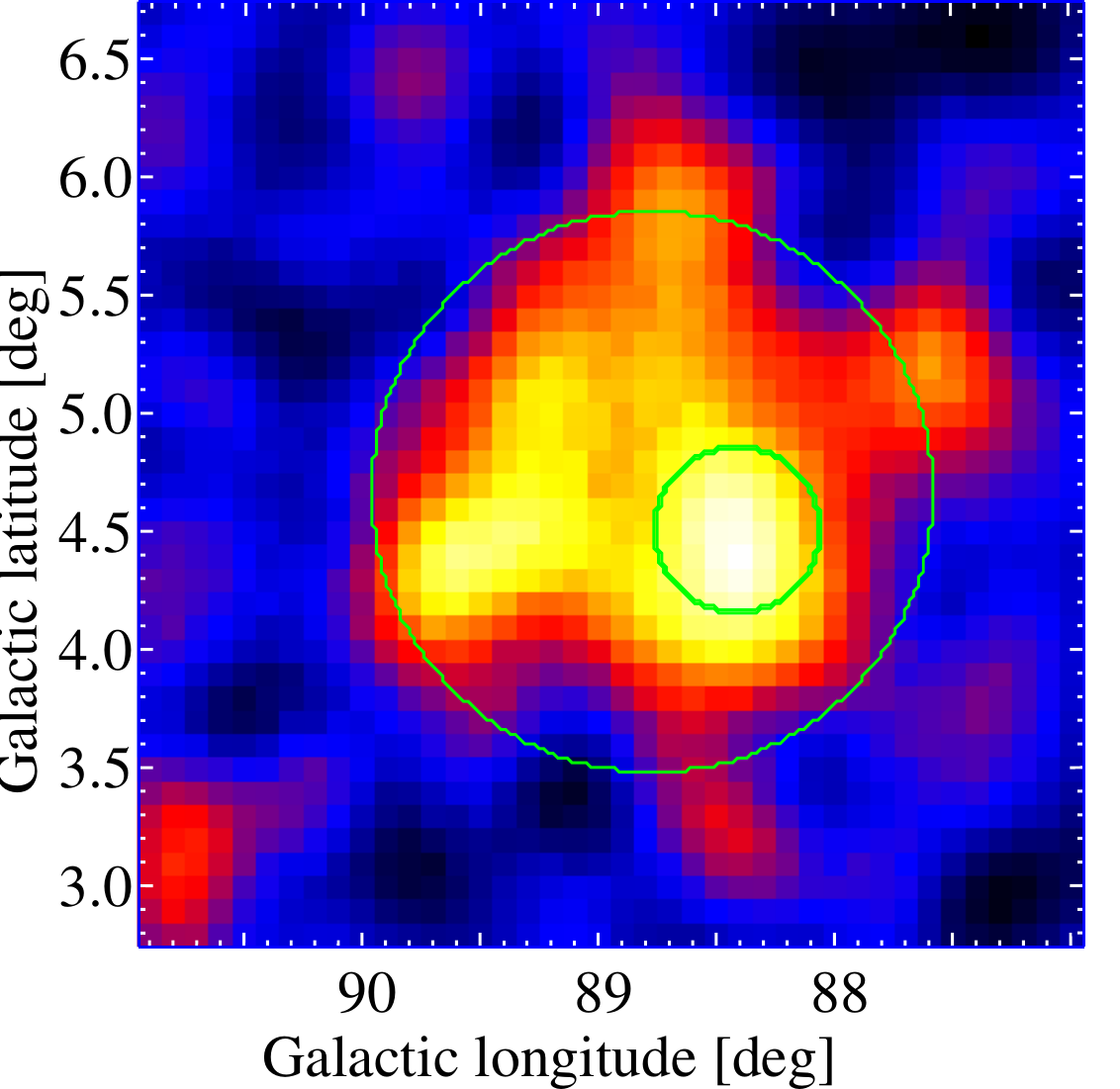}
\put(-120,120){\hbox{\large{\texttt{\color{white}{\textbf{(b)}}}}}}
\put(-60,70){\hbox{\tiny{\texttt{\color{black}{\textbf{shocked}}}}}}
\put(-52,65){\hbox{\tiny{\texttt{\color{black}{\textbf{CO}}}}}}
\\
\includegraphics[width=0.3\textwidth]{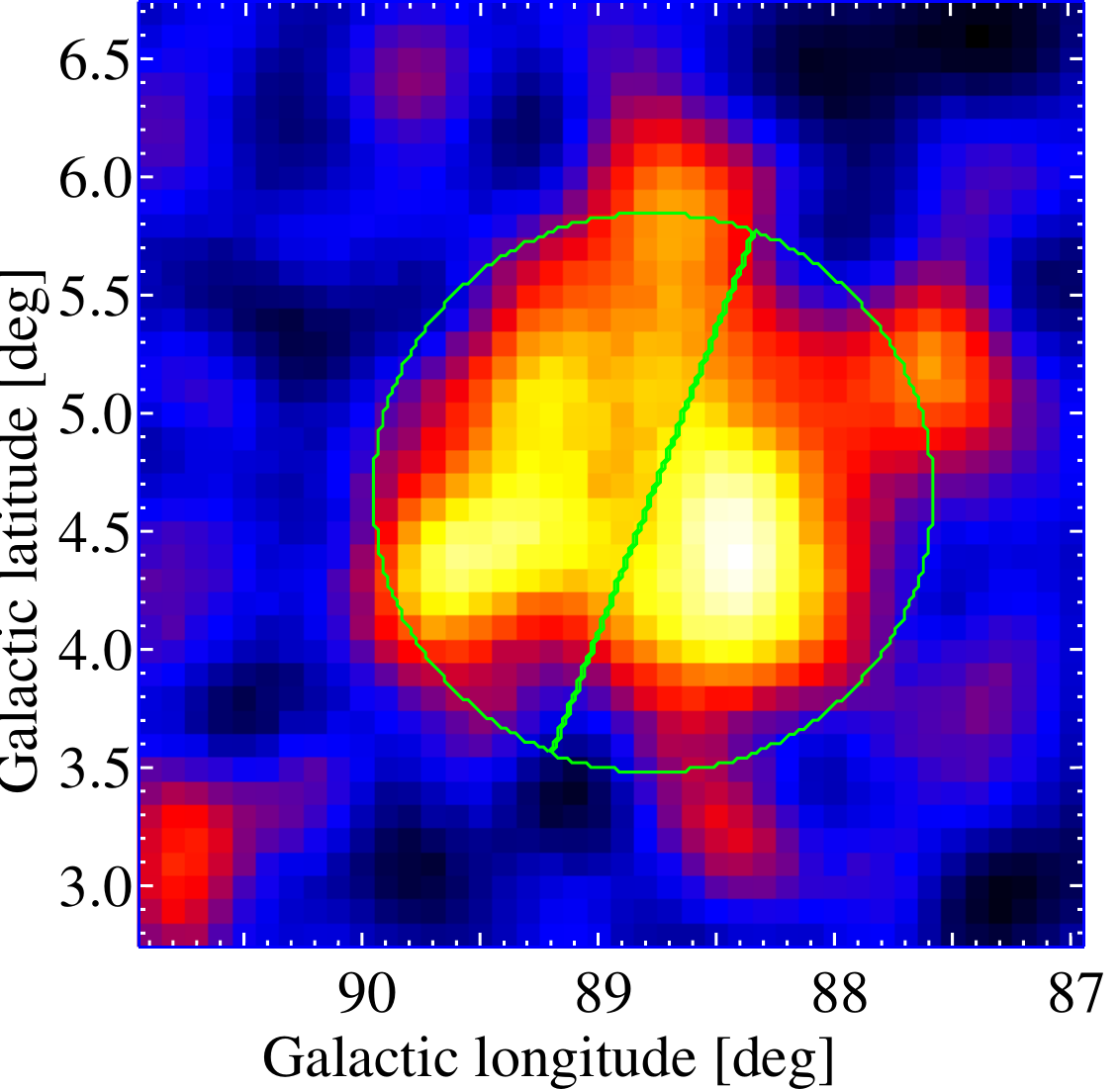}
\put(-120,120){\hbox{\large{\texttt{\color{white}{\textbf{(c)}}}}}} 
\put(-80,75){\hbox{\scriptsize{\texttt{\color{black}{\textbf{(1)}}}}}}
\put(-55,65){\hbox{\scriptsize{\texttt{\color{black}{\textbf{(2)}}}}}}
&
 \includegraphics[width=0.3\textwidth]{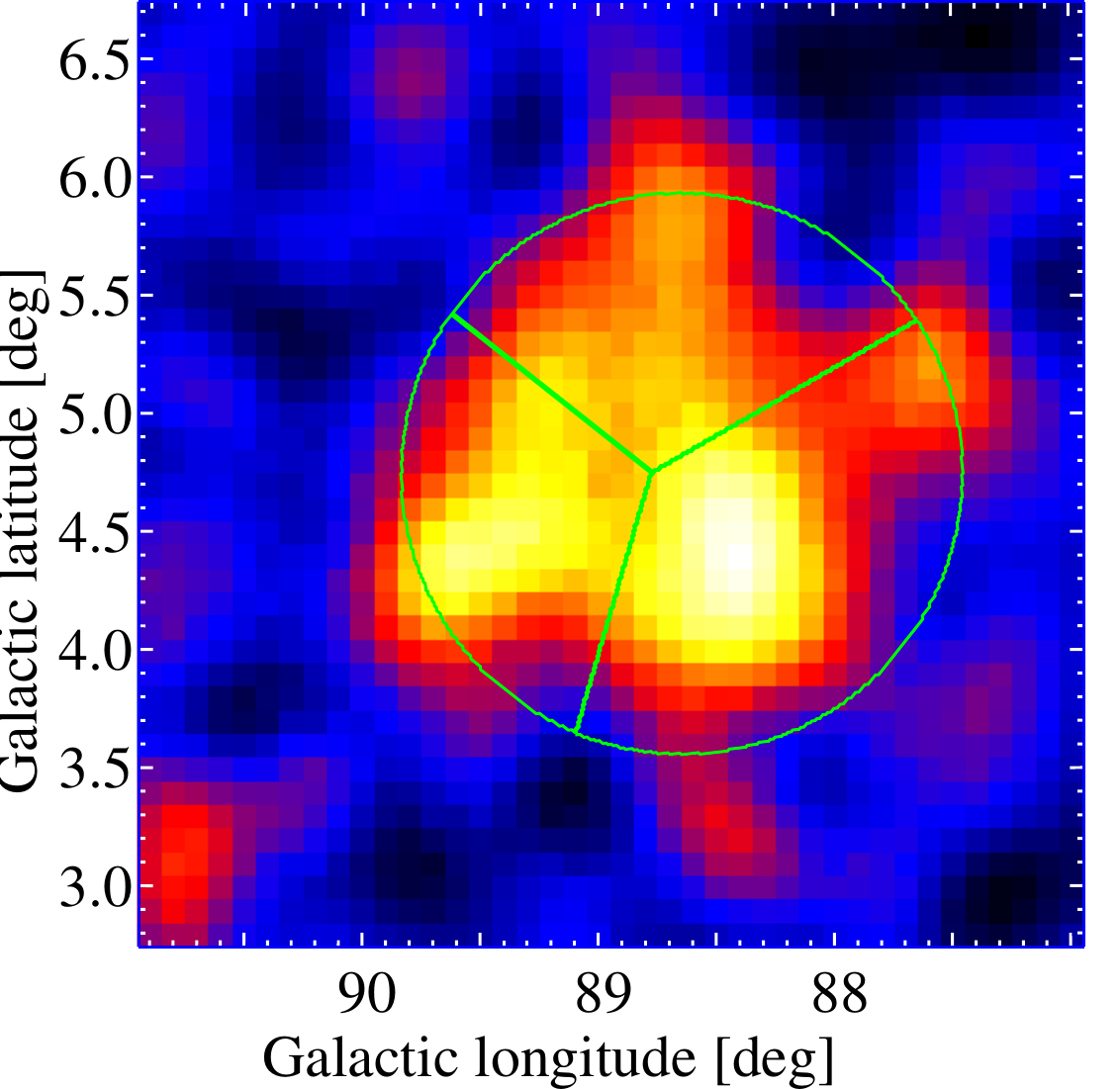}
\put(-120,120){\hbox{\large{\texttt{\color{white}{\textbf{(d)}}}}}}
\put(-80,65){\hbox{\scriptsize{\texttt{\color{black}{\textbf{(1)}}}}}}
\put(-65,95){\hbox{\scriptsize{\texttt{\color{black}{\textbf{(2)}}}}}}
\put(-55,65){\hbox{\scriptsize{\texttt{\color{black}{\textbf{(3)}}}}}}
\\
\includegraphics[width=0.38\textwidth]{zoom_or.pdf} & \\
\end{tabular}
 \caption{Emission associated with HB~21 as in Fig.~\ref{maps}c overlaid with
the boundaries of the templates used in the search for spectral variations
across the remnant described in~\ref{specvar}.}
 \label{split}
 \end{figure}

HB~21 is an extended source, located in an area rich in interstellar matter. The
interaction with nearby molecular clouds \citep{Byu06} might cause spectral
variation across the SNR, as suggested in the analysis by \citet{reichardt12}.
To further investigate the possibility, we repeat the analysis described in
\S\ref{morphology} after splitting the best-fit disk in several
different ways (Figure \ref{split}):
\begin{itemize}
 \item[a.] radio-emitting area (modeled with a uniform disk centered at
l=89\ddeg0 b=+4\ddeg7 with radius 1$^\circ$) and the remainder of the best-fit
disk  (Figure \ref{split}a);
 \item[b.] southern shocked-CO region (modeled as a uniform disk centered at
l=88\ddeg38
b=4\ddeg50 and radius 0\ddeg35) and the remainder of the best-fit disk (Figure
\ref{split}b);
 \item[c.] disk split into two halves to separate the southern shocked CO area
from the
rest (Figure \ref{split}c);
 \item[d.] finally, we test the division proposed by \citet{reichardt12}
splitting the uniform disk into three subregions covering $120^\circ$ each
(Figure \ref{split}d).
\end{itemize}

We first perform the analysis above 1 GeV to profit from the narrower PSF. For
this configuration we modeled the source
spectrum as a simple power law. The results are reported in
Table~\ref{spectrum_1gev}. There are no significant improvements in the
likelihood from splitting the disk in subregions ($< 2 \sigma$). 
This means that there is no evidence of spectral variations across the \g-ray
emitting region. Indeed, the spectral indices of the subregions are all
compatible
within statistical uncertainties.

\begin{deluxetable}{cccccc}
\tablecaption{Likelihood values and spectral indices for the models described in \S \ref{specvar},
splitting the single best--fit disk in different subregions. For each case we report the number of
additional degrees of freedom (d.o.f.) and the spectral indices.  Analysis
performed above 1~GeV. }
\tablenum{4}
\tablewidth{0pt}
\tablehead{\colhead{region splitting} & \colhead{$\Delta TS$} & \colhead{$\Delta$ d.o.f.} & \colhead{index 1} & \colhead{index 2} & \colhead{index 3}\\
\colhead{} & \colhead{} & \colhead{}  & \colhead{} & \colhead{} & \colhead{}}
\startdata
 single disk & 0 & 0 & 3.09$\pm$0.12  &  &\\
 a & 2.8 & 2 & 2.98$\pm$0.12 & 3.35$\pm$0.10 & \\
 b & 4.2 & 2  & 2.88$\pm$0.31 & 3.14$\pm$0.15  & \\
 c & 2.2 & 2  & 2.86$\pm$0.4 & 3.29$\pm$0.1 & \\
 d & 1.6 & 4 & 2.92$\pm$0.22 & 3.29$\pm$0.21 & 2.95$\pm$0.17 \\
 \enddata
\label{spectrum_1gev}
\end{deluxetable}

Then we perform the analysis over the full energy range above 100 MeV.
We fit the
spectrum of each subregion with a log-parabola function as for the uniform
disk. The results are reported
in  Table \ref{spectrum_100mev}. Even for this larger energy range we
find only marginal evidence ($<3\sigma$) of spectral
variations across the remnant and the values for the spectral parameters are
compatible within statistical errors.

\begin{deluxetable}{ccccc}
\tablecaption{Likelihood values and spectral indices for the models described in \S \ref{specvar},
splitting the single best--fit disk in different subregions. For each case we report the number of
additional degrees of freedom (d.o.f.) and the spectral indices.  The analysis is performed 
above 100~MeV. }
\tablenum{5}
\tablewidth{0pt}
\tablehead{\colhead{region splitting} & \colhead{$\Delta TS$} & \colhead{$\Delta$ d.o.f.} & \colhead{$\alpha$} & \colhead{$\beta$}\\
\colhead{} & \colhead{} & \colhead{}  & \colhead{} & \colhead{}}
\startdata
 single disk & 0 & 0 & 2.54$\pm$0.05  & 0.39$\pm$0.04 \\
 a & 13 & 3 & $\alpha_{radio}$=2.41$\pm$0.09 & $\beta_{radio}$=0.37$\pm$0.06 \\
    &  &   & $\alpha_{other}$=3.48$\pm$0.89 & $\beta_{other}$=0.74$\pm$0.39 \\
 b & 13.4 & 3 & $\alpha_{CO}$=2.27$\pm$0.06 & $\beta_{CO}$=0.38$\pm$0.04 \\
    &  &   & $\alpha_{other}$=2.61$\pm$0.21 & $\beta_{other}$=0.43$\pm$0.03 \\
 c & 8 & 3 & $\alpha_1$=2.47$\pm$0.06 & $\beta_1$=0.36$\pm$0.03 \\
    &  &   & $\alpha_2$=2.64$\pm$0.09 & $\beta_2$=0.45$\pm$0.03 \\
 d & 3 & 6 & $\alpha_1$=2.35$\pm$0.15 & $\beta_1$=0.38$\pm$0.10 \\
   &  &   & $\alpha_2$=2.75$\pm$0.29 & $\beta_2$=0.39$\pm$0.17 \\
   &  &   & $\alpha_3$=2.53$\pm$0.18 & $\beta_3$=0.43$\pm$0.13 \\
 \enddata
\label{spectrum_100mev}
\end{deluxetable}

In conclusion, we do not confirm
the claim by \citet{reichardt12} of a softer spectrum toward the clouds NW
and A. Spectral variations are not signifcant ($<3\sigma$), even
considering statistical uncertainties only. More data and a better handling of
the current dominating uncertainties related to the modeling of interstellar
emission (that were neglected in the study of \citealt{reichardt12}) are
required in order to further probe for possible \g-ray spectral variations
across HB~21. 

\section{$WMAP$ observations and radio spectrum}
\label{sec:radio}

The radio morphology of HB~21 is that of a large oblate shell (see contours in Figure \ref{split}a).
It has both a large angular diameter and high radio flux density, which have made it a favorable 
target for telescopes across the radio spectrum. High-resolution imaging reveals bright filaments 
and indentations along the periphery \citep{leahy98}. We use the 7-year all-sky data of
the {\it Wilkinson Microwave Anisotropy Probe} ({\it WMAP}) to extend the radio spectrum
of HB~21 above 10 GHz. Five bands are analyzed with effective central
frequencies ($\nu_{\rm eff}$) of 23 to 93 GHz \citep{Jarosik2011}.

The {\it WMAP} data are fit within a 4\degr\ square region
with a spatial template plus a sloping planar baseline
\citep[following][]{hewitt12}. For the
spatial template we use the map of HB~21 from the Sino-German 6-cm (4.8 GHz)
survey \citep[resolution of 9.5\arcmin ]{gao11}. 
We include a separate, freely-normalized point source to account for the bright nearby
extragalactic point source 3C~418 \citep{hill74}. The templates are smoothed to
the {\it WMAP} beam at each band \citep{weiland11} and fit to the data. 

\begin{figure}[htbp]
 \centering
\includegraphics[width=\textwidth]{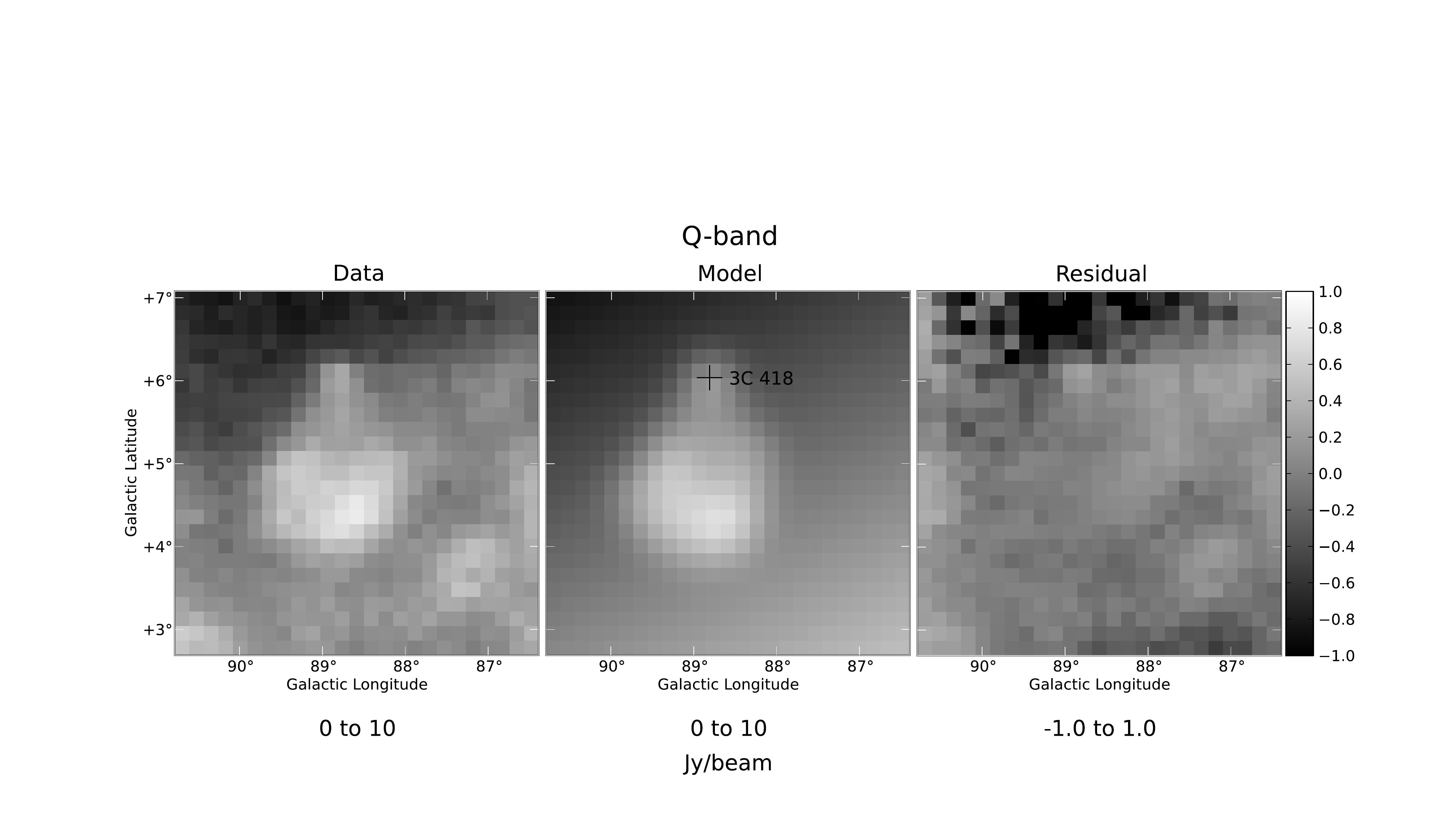}
 \caption{
Example of the template fit to the {\it WMAP} data within a 4\degr\ square
region of HB~21 in the Q band (61 GHz). The three panels present the 7-year
skymap image (left), the model resulting from a fit of the radio template plus
a sloping planar baseline and a point source at the position of 3C~418 marked by a cross
(center), and the fractional residuals (defined as the fit residual divided by
the model). The upper and lower limits of the linear color bar are given beneath
each image; data and model are on the same scale.
}
\label{fig:wmapfit} 
\end{figure}

Figure \ref{fig:wmapfit} shows an example of the {\it WMAP} skymap,
best-fit model and residual map for the Q band (61 GHz). 
Table \ref{tbl:wmap} lists the beam sizes and fitted fluxes with
errors estimated from the RMS of the fit residuals. 
To ensure that fluxes we extracted from the {\it WMAP} data are properly
calibrated, we also analyzed the bright radio sources
3C~58 and Cassiopeia~A using the same procedure and find agreement with \citet{weiland11}.
We also note that the addition of a template for 3C 418 does not produce a significant change in the measured fluxes of HB~21.

\begin{deluxetable}{cccc}
\tablecaption{$WMAP$ Flux Density for HB~21. Five bands are analyzed with effective central frequencies ($\nu_{\rm eff}$) of 23 to 93 GHz (see \S\ref{sec:radio}). \label{tbl:wmap}}
\tablewidth{0pt} 
\tablenum{6}
\tablehead{ \colhead{Band} & \colhead{$\nu_{eff}$} & \colhead{beam FWHM} &
\colhead{Flux Density}  \\ & \colhead{(GHz)} &
\colhead{(\degr)} & \colhead{(Jy)} }
\startdata
K	&22.7	& 0.93	&34$\pm$3  \\
Ka	&33.0	& 0.68 	&24$\pm$4 \\
Q	&40.6	& 0.53 	&20$\pm$5 \\
V	&60.5	& 0.35 	&17$\pm$8 \\
W	&93.0	& 0.23 	&56\tablenotemark{a} 
\enddata
\tablenotetext{a}{2$\sigma$ upper limit.} 
\end{deluxetable}

\begin{figure}[htbp]
 \centering
 \includegraphics[width=4.5in]{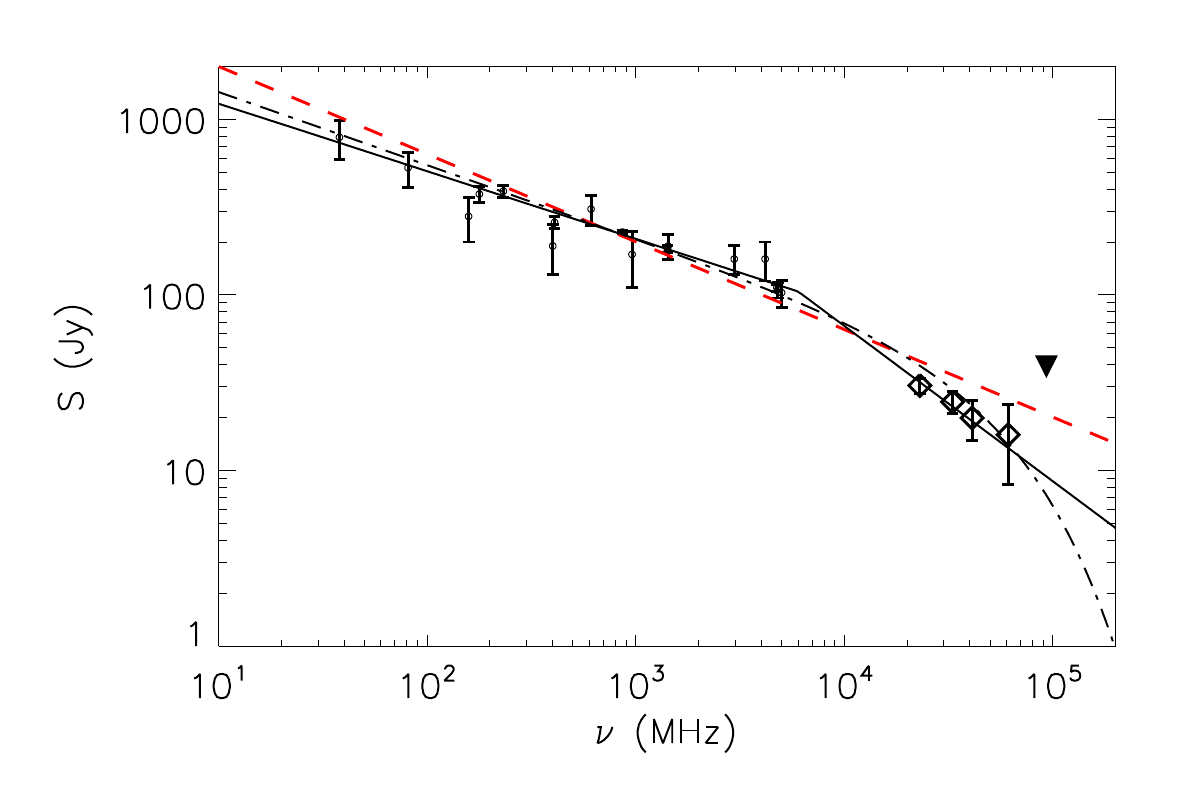}
 \caption{Integrated radio flux density of HB~21 as a function of frequency.
Data points below 10 GHz, circles, are from the literature (see Kothes, et al.
2006, and references therein).
The new {\it WMAP} data points are shown as open diamonds and presented in Table
\ref{tbl:wmap}. The upper limit is shown as a downward filled triangle. The
dashed red line shows a power-law fit to the entire radio spectrum. The solid
black line shows a fit to the data assuming a spectral break as described in
the text. The dot-dashed black line shows the fit assuming an exponential cutoff.}
\label{fig:radiosed} 
\end{figure}

The global radio spectrum of HB~21 from 38 MHz to 93 GHz is shown in Figure
\ref{fig:radiosed}. We include our {\it WMAP} analysis along with published flux
densities from the literature \citep[and references therein]{kothes06}. 
It is evident that a spectral break is present in the spectrum at high
frequencies. We first fit the radio spectrum with a single power law, S($\nu$) = S$_{0}$
$\nu^{-\alpha}$ and find a flux normalization at 1 GHz of S$_{0}$ = 201$\pm$3 Jy and
an index $\alpha$ = 0.50 $\pm$ 0.02. The fitted index is significantly steeper than
$\alpha$ = 0.38 determined from fitting only the data below 10 GHz because it does not
account for the break at high frequencies. 

We therefore include a spectral break of $\Delta\alpha$ = 0.5 at a frequency $\nu_{\rm b}$.
This is appropriate for synchrotron losses in a homogeneous source
of continuously injected electrons, as expected for middle-aged SNRs
 \citep[][Section~2.1]{leahy98,rey09}. 
With one additional free parameter,
we can now produce a good fit to the radio spectrum. We find the spectral break 
to have a significance of 5.3$\sigma$ by applying the F-test to compare the
$\chi^{2}$ fit to that of the simple power-law. The best fit parameters for the
radio spectrum of HB~21 are $\alpha$ = 0.38 $\pm$ 0.02 and $\nu_{\rm b}$ = 5.9 $\pm$ 1.2
GHz. Additionally, we tried to fit the spectrum using a power law with an exponential cutoff of the form,
S($\nu$) = S$_{0}$ $\nu^{-\alpha}$ exp($-\nu/\nu_{\rm c}$). This produces nearly as good a fit, with
$\alpha$ = 0.34 $\pm$ 0.02,  $\nu_{\rm c}$ = 29 $\pm$ 4 GHz, and a significance of 5.2$\sigma$.

The radio index, $\alpha$, is related to the particle index, $\Gamma_p$, by
$\Gamma_p$ = 2$\alpha$+1. The observed radio index below the break $\alpha$ = 0.38
gives $\Gamma_p$ $\sim$ 1.8,  which is similar to the spectral index obtained with the
\emph{Fermi}~LAT data in \S\ref{spectral}. We explore physical mechanisms which
could explain both the \g-ray and radio spectra in the
discussion in \S\ref{discussion}.

The observed break in the high-frequency spectrum cannot be explained by spectral 
variations across HB~21. \citet{leahy06} studied spectral variations using radio
observations at 408 and 1420 MHz. Manual fits to 36 individual regions across the
SNR show variations between 0.2--0.8, with a mean spectral index of 0.45 and a
standard deviation of 0.16. The brightest regions tend to have an index that is
flatter than average and than the canonical spectral index (0.5) from diffusive
shock acceleration. Thus, fitting the radio spectral index of the flux from the
entire remnant leads to an even flatter global index of 0.38. 

\section{Discussion} \label{discussion}
\subsection{Non-thermal Modeling} \label{models}

We have identified spatially extended \g-ray emission coincident with SNR HB~21,
indicating the presence of relativistic particles.  Determining the mechanism
responsible for \g-ray emission is crucial in order to measure the underlying
particle population accelerated by the SNR. To do so, we model emission from the
remnant using \texttt{isis}, the {\it Interactive Spectral Interpretation System}
\citep{isis}. For arbitrary particle momentum distributions, a suite of interaction models available in the literature is used to calculate the non-thermal emission spectrum \citep[and references therein]{houck2006} which is then fit to radio and \g-ray data. We used an updated interaction cross section for hadronic interactions \citep{Karlsson2008}.
The included prescriptions are in agreement with the results from \cite{sturner97}.

To constrain the emitting particle distribution we simultaneously fit radio and \g-ray
emission from non-thermal electrons and protons. Initially, we adopt the
simplifying assumption that all emission originates from a region characterized
by a constant matter density and magnetic field strength. This single emitting
zone is assumed to be equal to the size of the remnant derived from the best-fit
\g-ray disk. Additionally, the population of accelerated nuclei and electrons
is assumed to be described by the same particle distribution, here assumed to
follow a power-law with an exponential cutoff of the form dN/d$p$ $\propto$
$\eta_{\rm e, p}$ $p^{-\Gamma_p}$ $\times$ exp(--$p/p_{\rm max}$), where
$\eta_{\rm e}$/$\eta_{\rm p}$ gives the ratio of electrons to protons and $p$ 
the momentum. The normalization and maximum energy cutoff are left as free parameters, and
adjusted to fit the data. An exponential cutoff in the momentum 
spectra of electrons is expected when energy losses exceed the rate of 
energy gain from shock acceleration \citep{webb84}. In practice, we find that we
do not have sufficient spectral coverage to differentiate strongly between
an exponential cutoff or a broken power law.

To determine the total energy in relativistic particles from the normalization of
the non-thermal emission we must consider the physical conditions of the emitting
region. HB~21 is known to lie in the vicinity of several large molecular clouds.
The presence of optical $[$\ion{S}{2}$]$ but not oxygen line emission indicates slow
shocks $<$100 \kms\ into ambient densities of at least 2.5 cm$^{-3}$
\citep{mavromatakis07}. Shocked CO filaments are observed with densities of order
$\sim$10$^{2}$--10$^{4}$ cm$^{-3}$ and small filling factors $\leq 0.1$
\citep{koo01}. For neutral gas, the mean ambient HI density for the expanding
shell is $\sim$8 cm$^{-3}$ at a distance of 1.7 kpc \citep{koo91}.
\citet{reichardt12} estimated the molecular mass within HB~21 by integrating all
CO line emission between the velocity range from +0 to --20 \kms. This estimate
likely includes molecular gas outside the SNR, but is a conservative upper limit.
Adopting a distance of 1.7 kpc, the maximum molecular mass is
$5.5\times10^4$~\msol, the diameter of the SNR is 55 pc, and the maximum
volume-averaged molecular gas density is 25 cm$^{-3}$. Therefore, we assume a gas
density of 15 cm$^{-3}$, noting that this is uncertain by a factor of a few.

There are three primary emission mechanisms to produce \g-ray emission in SNRs. In
the so-called hadronic scenario the emission is dominated by \g-rays
radiated through the decay of $\pi^0$ mesons produced in collisions between
accelerated nuclei with the ambient gas. In the
leptonic scenarios
\g-ray emission results either from IC scattering of
relativistic electrons on low-energy photon
fields such as the cosmic microwave background (CMB), or
non-thermal bremsstrahlung.
As the matter density is increased, the bremsstrahlung contribution will rise and
dominate over the IC at densities $\gtrsim$1 cm$^{-3}$, unless the photon field is
greatly amplified above the CMB. HB~21 appears to have a high enough gas density 
that bremsstrahlung is expected to dominate over IC emission.
In modeling the non-thermal spectrum of HB~21 we
explored models in which the assumed physical conditions are modified such that
 each of these three emission mechanisms is dominant.

\begin{figure}[htbp]
\centering
\includegraphics[height=2.3in]{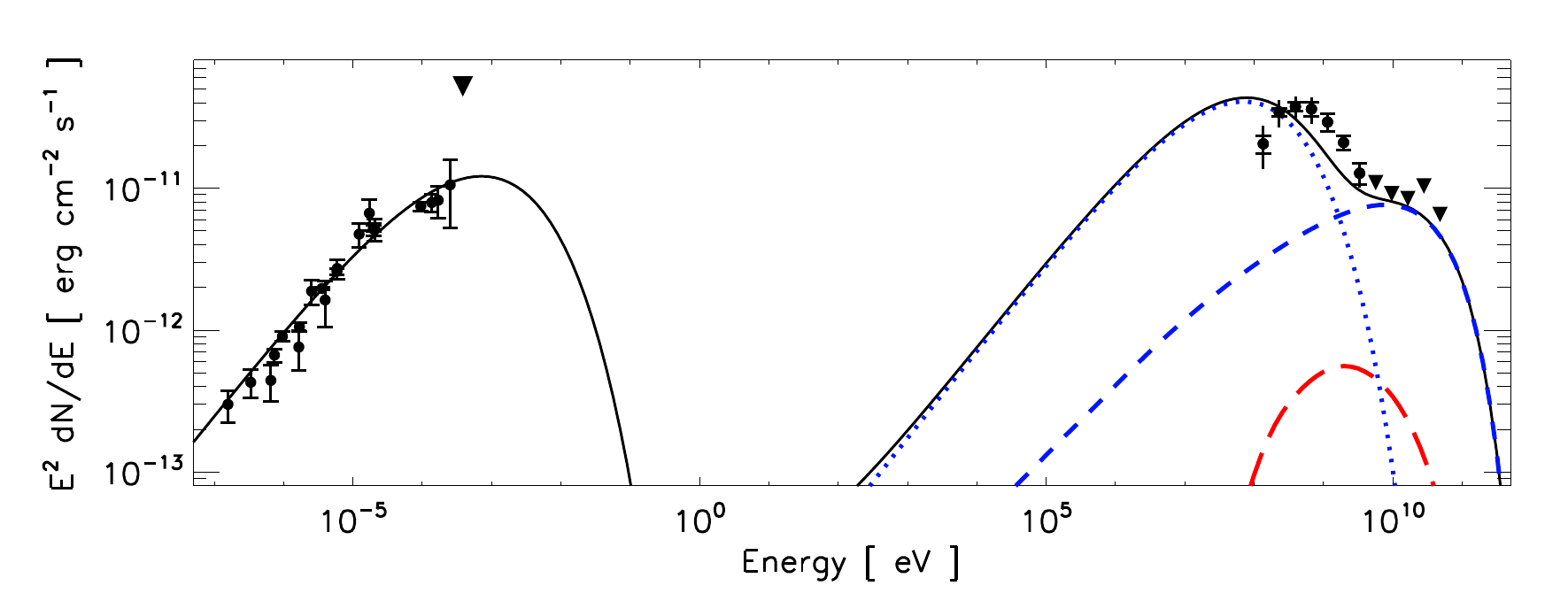} 
\put(-240,120){\hbox{\large{\texttt{\color{black}{\textbf{Inverse Compton}}}}}}\\
\includegraphics[height=2.3in]{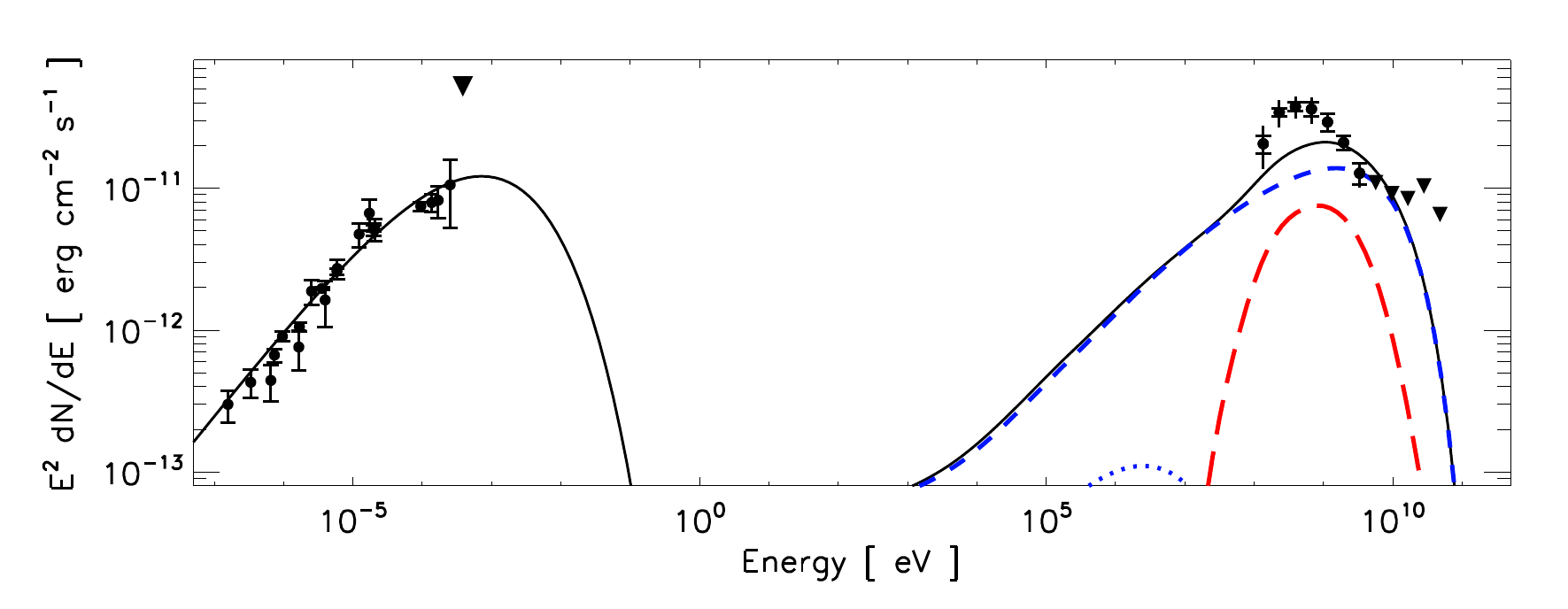} 
\put(-240,120){\hbox{\large{\texttt{\color{black}{\textbf{Bremsstrahlung}}}}}}\\
\includegraphics[height=2.3in]{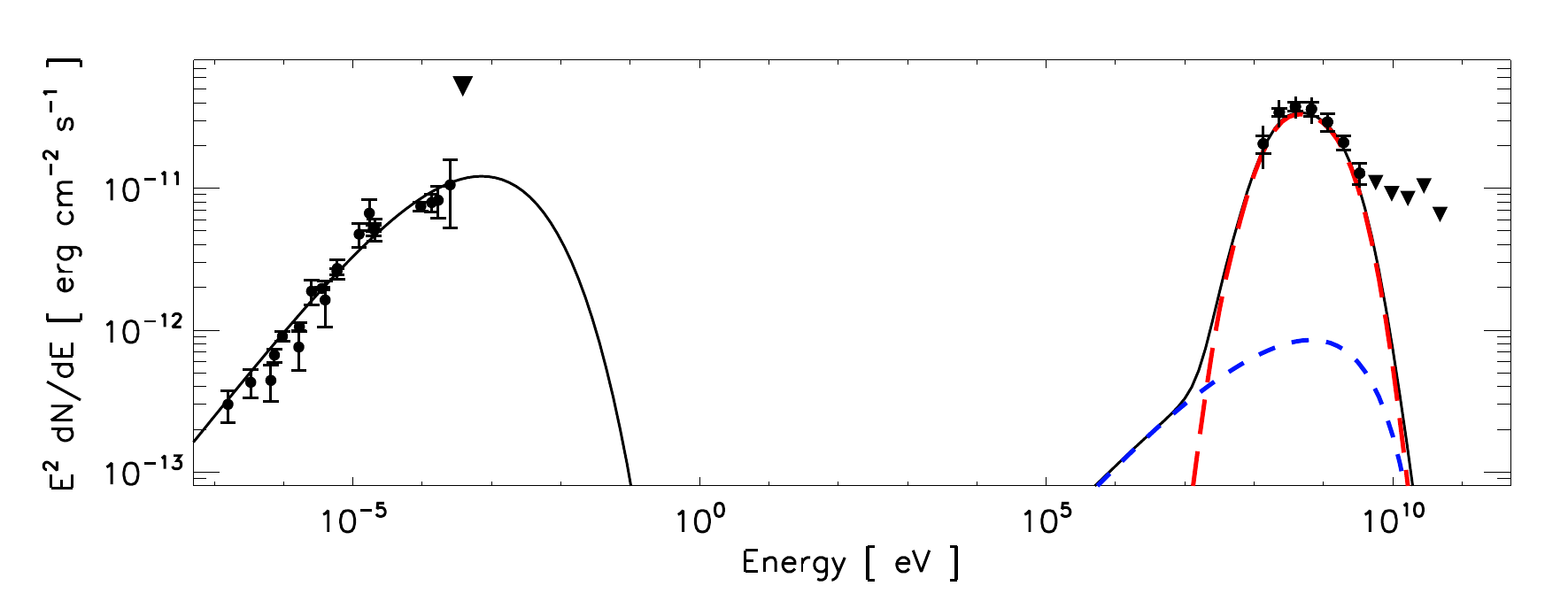} 
\put(-240,120){\hbox{\large{\texttt{\color{black}{\textbf{$\pi^0$-decay}}}}}}
\caption{Single-zone models for which IC (top), bremsstrahlung (middle) and $\pi^0$-decay (bottom) are the dominant emission mechanism (see Table~\ref{tbl:sed_models} for parameters). In each model the radio data are fit with a synchrotron component, shown as a black curve. The individual contributions of $\pi^0$-decay (long dashed), bremsstrahlung (short dashed), and IC emission from CMB (dotted) are shown. The sum of the \g-ray emission is shown by the solid curve. The leptonic and hadronic components are colored blue and red, respectively.}
\label{fig:sed_model}
\end{figure}

One-zone models for all three scenarios are presented as SED fits in Figure
\ref{fig:sed_model}. Parameters are given in Table \ref{tbl:sed_models}, including
the total energy of accelerated particles integrated above 1~GeV for protons, and
above 511~keV for electrons. We adjusted the ratio of electrons-to-protons to
differentiate between bremsstrahlung- and hadronic-dominated models. 
A ratio $\eta_{\rm e}$/$\eta_{\rm p}$ $\sim$ 0.01 is seen in cosmic rays
at Earth around 10 GeV \citep{gaisser90}, but an even lower ratio $\sim$0.001 may be 
expected from diffusive shock acceleration models \citep{rey08}. We cannot 
robustly constrain $\eta_{\rm e}$/$\eta_{\rm p}$ through our fits, so we choose 
characteristic values for each scenario.
While the chosen parameters are not unique in their ability to fit the broadband 
spectrum, they are representative.

Under our simple assumptions, a single zone with a single particle distribution
for both electrons and protons, we find that only hadronic models can reproduce
both the observed radio and \g-ray spectra. For $\eta_{\rm e}$/$\eta_{\rm
p}$ = 0.001 we find a momentum cutoff of 10 GeV/c and a magnetic field of $\sim$90
$\mu$G. Decreasing $\eta_{\rm e}$/$\eta_{\rm p}$ results in a higher magnetic
field strength to simultaneous fit the synchrotron normalization and break in
the radio. Bremsstrahlung dominates over neutral pion decay when $\eta_{\rm
e}$/$\eta_{\rm p}$ $\geq$ 0.05, but there is no combination of magnetic field
strength and momentum cutoff that can simultaneously produce the observed SED from
one electron population, as can be seen in Figure \ref{fig:sed_model}. To produce
a model in which IC emission dominates, we must adopt a density of
$\sim$0.1 cm$^{-3}$, which is well below gas density estimates. Therefore an
IC-dominated model is unlikely, and furthermore, cannot produce a good fit to the
data. The energetics of our hadronic model indicate $\sim$3$\times$10$^{49}$ erg
in accelerated cosmic ray protons and nuclei, which is comparable to that
estimated for other old SNRs in a dense environment detected by {\it Fermi}.

The failure of bremsstrahlung-dominated models is largely due to an inability to fit both the
observed spectral breaks in the radio and \g-rays. To explain synchrotron emission
at a peak frequency $\nu$ from an electron in a magnetic field B, requires the
electron have an energy E = 14.7($\nu_{\rm GHz}$/B$_{\mu \rm G}$)$^{1/2}$ GeV
\citep{rey08}. For the observed radio break at $\sim$6 GHz and \g-ray break at
$\sim$1 GeV to be explained by the same electron population requires a magnetic
field of $\sim$1 mG, which is far in excess of the magnetic field expected for a
SNR in such an evolved stage (unless the density is very high $\ga$10$^4$
cm$^{-3}$). However, molecular clouds have a well-known multiphase structure, so the
density of HB~21 is unlikely to be uniform, and the one-zone approximation may be 
overly simplistic.

We therefore explored relaxing the single-zone assumption by
modeling the radio and \g-ray emission as dominated by distinct regions.
Allowing the normalization and spectrum of the radio emission to be separately fit 
from \g-rays is physically motivated. 
Radio emission is observed from dense
filaments, but globally may be dominated by diffuse gas that fills a larger volume. 
For SNRs W44 and IC 443 the observed proton index from $\pi^0$-decay emission is softer than 
the electron index inferred from the radio spectrum \citep{pionbump}.

In this two-zone scenario, we found that both bremsstrahlung- and hadronic-dominated models can fit the data.
The cutoff in the accelerated particle spectrum responsible for \g-rays need not 
match that responsible for the radio emission, due to the different physical conditions.
It is also possible that high-energy CR electrons may cool in the dense filaments formed by shock-interaction with molecular clumps, 
or that CR protons may have largely escaped from the SNR \citep{aharonian96,malkov11}. In the latter case, we would expect 
that nearby clouds could be illuminated by the escaping CRs, but the geometry of
the clouds in relation to HB~21 is not well known. While multi-zone models appear as 
feasible as single zone models, they are not well constrained due to the  poor spatial resolution of the data at \g-rays 
and high-frequency radio. The energy in CRs in bremsstrahlung-dominated
two-zone models is several times 10$^{48}$ ergs, with a comparable energy in
accelerated nuclei and leptons.

\begin{deluxetable}{lcccccccccc}
\tablecaption{One-Zone Model Parameters \label{tbl:sed_models}}
\tablewidth{0pt}
\tablenum{7}
\tablehead{ 
\colhead{Model} & \colhead{Index} & \colhead{$p_{\rm max}$} & \colhead{$n_{\rm H}$} &
\colhead{$B_{\rm tot}$} & \colhead{$\eta_e$/$\eta_p$} & \colhead{$W_{p}$} & \colhead{$W_{e}$}  \\
& & \colhead{[GeV/c]} & \colhead{[cm$^{-3}$]} & \colhead{[$\mu$G]} & & \colhead{[erg]} &
\colhead{[erg]}
}
\startdata
IC			&1.76 & 100 	&0.1  & 2    &1	 & 1.3$\times$10$^{50}$ 	&  2.1$\times$10$^{51}$ 	\\ 
Brems.		&1.76 & 19	&15	 & 24  &0.1 & 6.4$\times$10$^{48}$	&  $3.0\times$10$^{48}$	\\ 
$\pi^0$-decay	&1.76 & 8.1	&15	 & 140  &0.001 & 3.0$\times$10$^{49}$	&  1.1$\times$10$^{47}$	\\ 
\enddata
\end{deluxetable}

\subsection{HB~21 and the population of detected \g-ray SNRs}

Here we briefly discuss HB~21 in comparison to other \g-ray SNRs known to be interacting with molecular clouds.
The total luminosity of HB~21 above 100 MeV at a distance of 1.7 kpc is
$(3.3 \pm 0.6)\times$10$^{34}$ erg s$^{-1}$.
Other \g-ray-detected MM SNRs, such as W44 and IC 443, have luminosities of
$\sim$10$^{35}$ erg s$^{-1}$ and associated cloud masses of $>$10$^{4}$ M$_\odot$.
While the total CO line emission along the line of sight to HB~21 indicates
a large cloud mass, this could be due to the Cygnus OB7 complex, which
lies along the line of sight at a similar velocity to HB~21, but at a
distance of only 0.8 kpc, between us and the SNR. It is therefore possible that
the low luminosity of HB~21 is due to the SNR currently encountering only 
a relatively small reservoir of material.

Flat radio indices, as for HB~21, are observed for other \g-ray detected SNRs
known to be interacting with molecular
clouds,
such as IC 443 and W44.
\citet{leahy06}
proposed two mechanisms to produce the observed flat spectrum: ionization losses
due to emission from regions of high density, and low-frequency absorption by
thermal electrons in the post-shock cooling regions. Alternatively,
\citet{uchiyama10} proposed that re-acceleration takes place in the compressed
cloud, resulting in a hardening of the spectrum of existing accelerated
particles.
Though it is likely interacting with a less-dense environment than HB~21, the latter
model was shown to plausibly explain radio and \g-ray emission from SNR S147
\citep{katsuta12}.

HB~21 shows a cutoff or break in the \g-ray spectrum mirroring that of the
underlying particle
spectrum, which is typical of middle-aged SNRs
detected by the LAT (e.g. the Cygnus Loop, \citealt{cyg11}, and W28,
\citealt{w28ferm}). 
This is in agreement with the circumstantial evidence
for HB~21 itself to be a middle-aged/old remnant (see \S\ref{intro}). Given the
long timescales for radiative losses via proton-proton collision, bremsstrahlung or synchrotron losses for GeV particles,
it is unlikely that such spectral curvature is produced by radiative losses.
Several mechanisms have been proposed, including runaway CRs
illuminating nearby clouds \citep{gabici09}, the aforementioned re-acceleration
in
highly-compressed shocks at the interaction sites \citep{uchiyama10}, and
magnetic damping of Alfven waves in a partially ionized medium that leads to a
break in the particle spectrum \citep{malkov11}. All these mechanisms appear
viable for the case of HB~21, and could produce different particle distributions
for electrons and protons remaining in the SNR. 
Given the many shared similarities with other MM SNRs, HB~21 appears to be an extension 
of this \g-ray class to lower luminosities.

\section{Summary}
We analyzed \emph{Fermi}~LAT \g-ray data from the region around HB21, a
MM SNR. We detect significant \g-ray emission ($\sim$29$\sigma$)
associated with the
remnant.
The emission is best modeled by a disk centered at
(l,b) = (88\ddeg75 $\pm$ 0\ddeg04, +4\ddeg65 $\pm$ 0\ddeg06) with a radius r = 1\ddeg19 $\pm$ 0\ddeg06
with a systematic
uncertainty on the position of $\sim$0\ddeg25 and on the radius of
$\sim$0\ddeg24, so it is well-resolved by the LAT for energies
greater
than 1~GeV. The \g-ray emission extends over the whole area of the non-thermal radio shell, larger
than the X-ray emitting thermal core. The emission in \g-rays may extend beyond the
radio shell in a region rich of interstellar matter in the north
western part of the SNR. Furthermore, the brightest \g-ray emitting region coincides with known
shocked molecular clumps. Both results are suggestive that collisions of shock-accelerated particles
with interstellar matter are responsible for the observed \g-ray emission. No spectral variations
across the \g-ray emitting region that would further support this hypothesis were detected with the
current observations. 

The spectrum is best modeled by a curved function, indicative of a cutoff or
break in the spectrum of the accelerated particles, typical of middle-aged/old
SNRs in a dense interstellar environment. The total \g-ray luminosity of HB~21
above 100~MeV is estimated to be $(3.3\pm0.6)\times10^{34}$~erg~s$^{-1}$,
fainter than other SNRs interacting with molecular clouds detected by the LAT. This can be
explained by the lower mass of the molecular clouds supposed to be in
interaction with the remnant. 

\cite{reichardt12} recently reported a lower luminosity for HB~21.
We assume a distance of 1.7 kpc, based on the arguments of
\cite{Byu06}, while
their work assumes a nearer distance of 0.8 kpc adopted in numerous earlier
works. Taking this difference into account, our values of the 0.1-10 GeV flux
and luminosity of SNR HB~21 are in
agreement. Even
considering statistical uncertainties only, we do not find any
significant evidence for spectral variations across the SNR, as suggested for
cloud NW and A in
their work.

We complement the \g-ray analysis by exploiting  the {\it WMAP} 7-year observations
from 23 to 93 GHz, obtaining the first 
detections of HB~21 at these energies. By combining {\it WMAP} with lower-energies 
radio observations we find that the radio spectral index of HB~21 steepens significantly 
above 10~GHz. This spectral feature in the radio helps to constrain the relativistic electron 
spectrum and constrain possible physical parameters in simple non-thermal radiation models.

An IC origin of the \g-ray emission is 
disfavored because it would require unrealistically low interstellar densities
in order to 
prevent bremsstrahlung from dominating. $\pi^0$
decay due to nuclei interactions can reproduce the data well. 
Bremsstrahlung and synchrotron emission from a single population of energetic
electrons cannot reproduce both the \g-ray and radio SEDs.  Based on the most
likely values for the ISM densities over the volume of the remnant, in the
hadronic-dominated scenario accelerated nuclei contribute a total energy of
$\sim$3$\times$10$^{49}$ ergs. This is reduced to several times 10$^{48}$ ergs 
under a two-zone bremsstrahlung-dominated model, with a comparable energy in
leptonic CRs.
 
\smallskip
The \textit{Fermi}-LAT Collaboration acknowledges generous ongoing support from
a number of agencies and institutes that have supported both the development and
the operation of the LAT as well as scientific data analysis. These include the
National Aeronautics and Space Administration and the Department of Energy in
the United States, the Commissariat \`a l'Energie Atomique and the Centre
National de la Recherche Scientifique/Institut National de Physique Nucl\'eaire
et de Physique des Particules in France, the Agenzia Spaziale Italiana and the
Istituto Nazionale di Fisica Nucleare in Italy, the Ministry of Education,
Culture, Sports, Science and Technology (MEXT), High Energy Accelerator Research
Organization (KEK) and Japan Aerospace Exploration Agency (JAXA) in Japan, and
the K.~A.~Wallenberg Foundation, the Swedish Research Council and the Swedish
National Space Board in Sweden. 
Additional support for science analysis during the operations phase is
gratefully acknowledged from the Istituto Nazionale di Astrofisica in Italy and
the Centre National d'\'Etudes Spatiales in France.

The Lovell Telescope is owned and operated by the University of Manchester as part of the Jodrell Bank Centre for Astrophysics with support from the Science and Technology Facilities Council of the United Kingdom. 
The Westerbork Synthesis Radio Telescope is operated by Netherlands Foundation for Radio Astronomy, ASTRON.
\bibliographystyle{apj.bst}
\bibliography{./hb21_references}

\begin{thebibliography}{72}
\expandafter\ifx\csname natexlab\endcsname\relax\def\natexlab#1{#1}\fi

\bibitem[{{Abdo} {et~al.}(2009{\natexlab{a}}){Abdo}, {Ackermann}, {Ajello},
  {Baldini}, {Ballet}, {Barbiellini}, {Baring}, {Bastieri}, {Baughman},
  {Bechtol}, {Bellazzini}, {Berenji}, {Blandford}, {Bloom}, {Bonamente},
  {Borgland}, {Bouvier}, {Bregeon}, {Brez}, {Brigida}, {Bruel}, {Burnett},
  {Buson}, {Caliandro}, {Cameron}, {Caraveo}, {Casandjian}, {Cecchi}, {{\c
  C}elik}, {Chekhtman}, {Cheung}, {Chiang}, {Ciprini}, {Claus}, {Cohen-Tanugi},
  {Cominsky}, {Conrad}, {Cutini}, {Dermer}, {de Angelis}, {de Palma}, {Digel},
  {Dormody}, {Silva}, {Drell}, {Dubois}, {Dumora}, {Farnier}, {Favuzzi},
  {Fegan}, {Focke}, {Fortin}, {Frailis}, {Fukazawa}, {Funk}, {Fusco},
  {Gargano}, {Gasparrini}, {Gehrels}, {Germani}, {Giavitto}, {Giebels},
  {Giglietto}, {Giordano}, {Glanzman}, {Godfrey}, {Grenier}, {Grondin},
  {Grove}, {Guillemot}, {Guiriec}, {Hanabata}, {Harding}, {Hayashida}, {Hays},
  {Hughes}, {Jackson}, {J{\'o}hannesson}, {Johnson}, {Johnson}, {Johnson},
  {Kamae}, {Katagiri}, {Kataoka}, {Katsuta}, {Kawai}, {Kerr}, {Kn{\"o}dlseder},
  {Kocian}, {Kuss}, {Lande}, {Latronico}, {Lemoine-Goumard}, {Longo},
  {Loparco}, {Lott}, {Lovellette}, {Lubrano}, {Makeev}, {Mazziotta}, {McEnery},
  {Meurer}, {Michelson}, {Mitthumsiri}, {Mizuno}, {Moiseev}, {Monte},
  {Monzani}, {Morselli}, {Moskalenko}, {Murgia}, {Nakamori}, {Nolan}, {Norris},
  {Nuss}, {Ohsugi}, {Okumura}, {Omodei}, {Orlando}, {Ormes}, {Paneque},
  {Parent}, {Pelassa}, {Pepe}, {Pesce-Rollins}, {Piron}, {Porter}, {Rain{\`o}},
  {Rando}, {Razzano}, {Reimer}, {Reimer}, {Reposeur}, {Ritz}, {Rodriguez},
  {Romani}, {Roth}, {Ryde}, {Sadrozinski}, {Sanchez}, {Sander}, {Saz
  Parkinson}, {Scargle}, {Schalk}, {Sgr{\`o}}, {Siskind}, {Smith}, {Smith},
  {Spandre}, {Spinelli}, {Strickman}, {Suson}, {Tajima}, {Takahashi},
  {Takahashi}, {Tanaka}, {Thayer}, {Thayer}, {Thompson}, {Tibaldo}, {Tibolla},
  {Torres}, {Tosti}, {Tramacere}, {Uchiyama}, {Usher}, {Vasileiou}, {Venter},
  {Vilchez}, {Vitale}, {Waite}, {Wang}, {Winer}, {Wood}, {Yamazaki}, {Ylinen},
  \& {Ziegler}}]{w51cfermi}
{Abdo}, A.~A., {et~al.} 2009{\natexlab{a}}, \apjl, 706, L1

\bibitem[{{Abdo} {et~al.}(2009{\natexlab{b}}){Abdo}, {Ackermann}, {Ajello},
  {Ampe}, {Anderson}, {Atwood}, {Axelsson}, {Bagagli}, {Baldini}, {Ballet}, \&
  et~al.}]{abd09}
---. 2009{\natexlab{b}}, Astroparticle Physics, 32, 193

\bibitem[{{Abdo} {et~al.}(2010{\natexlab{a}}){Abdo}, {Ackermann}, {Ajello},
  {Allafort}, {Baldini}, {Ballet}, {Barbiellini}, {Bastieri}, {Bechtol},
  {Bellazzini}, {Berenji}, {Blandford}, {Bloom}, {Bonamente}, {Borgland},
  {Bouvier}, {Brandt}, {Bregeon}, {Brigida}, {Bruel}, {Buehler}, {Buson},
  {Caliandro}, {Cameron}, {Caraveo}, {Carrigan}, {Casandjian}, {Cecchi}, {{\c
  C}elik}, {Chekhtman}, {Chiang}, {Ciprini}, {Claus}, {Cohen-Tanugi}, {Conrad},
  {Dermer}, {de Palma}, {Silva}, {Drell}, {Dubois}, {Dumora}, {Farnier},
  {Favuzzi}, {Fegan}, {Fukazawa}, {Fukui}, {Funk}, {Fusco}, {Gargano},
  {Gehrels}, {Germani}, {Giglietto}, { Giordano}, {Glanzman}, {Godfrey},
  {Grenier}, {Grove}, {Guiriec}, {Hadasch}, {Hanabata}, {Harding}, {Hays},
  {Horan}, {Hughes}, {J{\'o}hannesson}, {Johnson}, {Johnson}, {Kamae},
  {Katagiri}, {Kataoka}, {Kn{\"o}dlseder}, {Kuss}, {Lande}, {Latronico}, {Lee},
  {Lemoine-Goumard}, {Llena Garde}, {Longo}, {Loparco}, {Lovellette},
  {Lubrano}, {Makeev}, {Mazziotta}, {Michelson}, {Mitthumsiri}, {Mizuno},
  {Moiseev}, {Monte}, {Monzani}, {Morselli}, {Moskalenko}, {Murgia},
  {Nakamori}, {Nolan}, {Norris}, {Nuss}, {Ohno}, {Ohsugi}, {Omodei}, {Orlando},
  {Ormes}, {Ozaki}, {Panetta}, {Parent}, {Pelassa}, {Pepe}, {Pesce-Rollins},
  {Piron}, {Porter}, {Rain{\`o}}, {Rando}, {Razzano}, {Reimer}, {Reimer},
  {Reposeur}, {Rodriguez}, {Roth}, {Sadrozinski}, {Sander}, {Saz Parkinson},
  {Sgr{\`o}}, {Siskind}, {Smith}, {Smith}, {Spandre}, {Spinelli}, {Strickman},
  {Suson}, {Tajima}, {Takahashi}, {Takahashi}, {Tanaka}, {Thayer}, {Thayer},
  {Thompson}, {Tibaldo}, {Tibolla}, {Torres}, {Tosti}, {Uchiyama}, {Uehara},
  {Usher}, {Vasileiou}, {Vilchez}, {Vitale}, {Waite}, {Wang}, {Winer}, {Wood},
  {Yamamoto}, {Yamazaki}, {Yang}, {Ylinen}, \& {Ziegler}}]{w28ferm}
---. 2010{\natexlab{a}}, \apj, 718, 348

\bibitem[{{Abdo} {et~al.}(2010{\natexlab{b}}){Abdo}, {Ackermann}, {Ajello},
  {Baldini}, {Ballet}, {Barbiellini}, {Bastieri}, {Bechtol}, {Bellazzini},
  {Bloom}, {Bonamente}, {Borgland}, {Bouvier}, {Bregeon}, {Brez}, {Brigida},
  {Bruel}, {Buehler}, {Buson}, {Caliandro}, {Cameron}, {Caraveo}, {Casandjian},
  {Cecchi}, {{\c C}elik}, {Cheung}, {Chiang}, {Ciprini}, {Claus},
  {Cohen-Tanugi}, {Conrad}, {Dermer}, {de Palma}, {Digel}, {Silva}, {Drell},
  {Dumora}, {Favuzzi}, {Funk}, {Fusco}, {Gargano}, {Gehrels}, {Giglietto},
  {Giordano}, {Giroletti}, {Glanzman}, {Godfrey}, {Grenier}, {Grondin},
  {Grove}, {Guillemot}, {Guiriec}, { Hadasch}, {Hanabata}, {Harding},
  {Hayashida}, {Hays}, {Horan}, {Hughes}, {Jackson}, {J{\'o}hannesson},
  {Johnson}, {Johnson}, {Kamae}, {Katagiri}, {Kataoka}, {Katsuta},
  {Kn{\"o}dlseder}, {Kuss}, {Lande}, {Latronico}, {Lee}, {Lemoine-Goumard},
  {Longo}, {Loparco}, {Lovellette}, {Lubrano}, {Makeev}, {Mazziotta}, {Mizuno},
  {Monte}, {Monzani}, {Morselli}, {Moskalenko}, {Murgia}, {Naumann-Godo},
  {Nolan}, {Norris}, {Nuss}, {Ohsugi}, {Okumura}, {Omodei}, {Orlando}, {Ormes},
  {Pelassa}, {Pepe}, {Pesce-Rollins}, {Piron}, {Rain{\`o}}, {Rando}, {Razzano},
  {Reimer}, {Reimer}, {Reposeur}, {Ripken}, {Roth}, { Sadrozinski}, {Sander},
  {Saz Parkinson}, {Sgr{\`o}}, {Siskind}, {Smith}, {Smith}, {Spinelli},
  {Strickman}, {Suson}, {Tajima}, {Takahashi}, {Takahashi}, {Tanaka},
  {Tibaldo}, {Tibolla}, {Torres}, {Tosti}, {Tramacere}, {Uchiyama}, {Usher},
  {Vandenbroucke}, {Vasileiou}, {Vitale}, {Waite}, {Wang}, {Winer}, {Wood},
  {Ylinen}, \& {Ziegler}}]{w49bfermi}
---. 2010{\natexlab{b}}, \apj, 722, 1303

\bibitem[{{Abdo} {et~al.}(2010{\natexlab{c}}){Abdo}, {Ackermann}, {Ajello},
  {Baldini}, {Ballet}, {Barbiellini}, {Baring}, {Bastieri}, {Baughman},
  {Bechtol}, {Bellazzini}, {Berenji}, {Blandford}, {Bloom}, {Bonamente},
  {Borgland}, {Bregeon}, {Brez}, {Brigida}, {Bruel}, {Burnett}, {Buson},
  {Caliandro}, {Cameron}, {Caraveo}, {Casandjian}, {Cecchi}, {{\c C}elik},
  {Chekhtman}, {Cheung}, {Chiang}, {Ciprini}, {Claus}, {Cognard},
  {Cohen-Tanugi}, {Cominsky}, {Conrad}, {Cutini}, {Dermer}, {de Angelis}, {de
  Palma}, {Digel}, {do Couto e Silva}, {Drell}, {Dubois}, {Dumora}, {Espinoza},
  {Farnier}, {Favuzzi}, {Fegan}, {Focke}, { Fortin}, {Frailis}, {Fukazawa},
  {Funk}, {Fusco}, {Gargano}, {Gasparrini}, {Gehrels}, {Germani}, {Giavitto},
  {Giebels}, {Giglietto}, {Giordano}, {Glanzman}, {Godfrey}, {Grenier},
  {Grondin}, {Grove}, {Guillemot}, {Guiriec}, {Hanabata}, {Harding},
  {Hayashida}, {Hays}, {Hughes}, {Jackson}, {J{\'o}hannesson}, {Johnson},
  {Johnson}, {Johnson}, {Kamae}, {Katagiri}, {Kataoka}, {Katsuta}, {Kawai},
  {Kerr}, {Kn{\"o}dlseder}, {Kocian}, {Kramer}, {Kuss}, {Lande}, {Latronico},
  {Lemoine-Goumard}, {Longo}, {Loparco}, {Lott}, {Lovellette}, {Lubrano},
  {Lyne}, {Madejski}, {Makeev}, {Mazziotta}, {McEnery}, {Meurer}, {Michelson},
  {Mitthumsiri}, {Mizuno}, {Monte}, {Monzani}, {Morselli}, {Moskalenko},
  {Murgia}, {Nakamori}, {Nolan}, {Norris}, {Noutsos}, {Nuss}, {Ohsugi},
  {Omodei}, {Orlando}, {Ormes}, {Paneque}, {Parent}, {Pelassa}, {Pepe},
  {Pesce-Rollins}, {Piron}, {Porter}, {Rain{\`o}}, {Rando}, {Razzano},
  {Reimer}, {Reimer}, {Reposeur}, {Rochester}, {Rodriguez}, {Romani}, {Roth},
  {Ryde}, {Sadrozinski}, {Sanchez}, {Sander}, {Parkinson}, {Scargle},
  {Sgr{\`o}}, {Siskind}, {Smith}, {Smith}, {Spandre}, {Spinelli}, {Stappers},
  {Stecker}, {Strickman}, {Suson}, {Tajima}, {Takahashi}, {Takahashi}, {
  Tanaka}, {Thayer}, {Thayer}, {Theureau}, {Thompson}, {Tibaldo}, {Tibolla},
  {Torres}, {Tosti}, {Tramacere}, {Uchiyama}, {Usher}, {Vasileiou}, {Venter},
  {Vilchez}, {Vitale}, {Waite}, {Wang}, {Winer}, {Wood}, {Yamazaki}, {Ylinen},
  \& {Ziegler}}]{w44ferm}
---. 2010{\natexlab{c}}, Science, 327, 1103

\bibitem[{{Abdo} {et~al.}(2010{\natexlab{d}}){Abdo}, {Ackermann}, {Ajello},
  {Baldini}, {Ballet}, {Barbiellini}, {Bastieri}, {Baughman}, {Bechtol},
  {Bellazzini}, {Berenji}, {Blandford}, {Bloom}, {Bonamente}, {Borgland},
  {Bregeon}, {Brez}, {Brigida}, {Bruel}, {Burnett}, {Buson}, {Caliandro},
  {Cameron}, {Caraveo}, {Casandjian}, {Cecchi}, {{\c C}elik}, {Chekhtman},
  {Cheung}, {Chiang}, {Cillis}, {Ciprini}, {Claus}, {Cohen-Tanugi}, {Cominsky},
  {Conrad}, {Cutini}, {Dermer}, {de Angelis}, {de Palma}, {Silva}, {Drell},
  {Drlica-Wagner}, {Dubois}, {Dumora}, {Farnier}, {Favuzzi}, {Fegan}, {Focke},
  {Fortin}, {Frailis}, { Fukazawa}, {Funk}, {Fusco}, {Gargano}, {Gasparrini},
  {Gehrels}, {Germani}, {Giavitto}, {Giebels}, {Giglietto}, {Giordano},
  {Glanzman}, {Godfrey}, {Grenier}, {Grondin}, {Grove}, {Guillemot}, {Guiriec},
  {Hanabata}, {Harding}, {Hayashida}, {Hughes}, {Jackson}, {J{\'o}hannesson},
  {Johnson}, {Johnson}, {Johnson}, {Kamae}, {Katagiri}, {Kataoka}, {Kawai},
  {Kerr}, {Kn{\"o}dlseder}, {Kocian}, {Kuss}, {Lande}, {Latronico}, {Lee},
  {Lemoine-Goumard}, {Longo}, {Loparco}, {Lott}, {Lovellette}, {Lubrano},
  {Madejski}, {Makeev}, {Mazziotta}, {Meurer}, {Michelson}, {Mitthumsiri},
  {Moiseev}, {Monte}, {Monzani}, { Morselli}, {Moskalenko}, {Murgia},
  {Nakamori}, {Nolan}, {Norris}, {Nuss}, {Ohsugi}, {Orlando}, {Ormes}, {Ozaki},
  {Paneque}, {Panetta}, {Parent}, {Pelassa}, {Pepe}, {Pesce-Rollins}, {Piron},
  {Porter}, {Rain{\`o}}, {Rando}, {Razzano}, {Reimer}, {Reimer}, {Reposeur},
  {Rochester}, {Rodriguez}, {Romani}, {Roth}, {Ryde}, {Sadrozinski}, {Sanchez},
  {Sander}, {Saz Parkinson}, {Scargle}, {Sgr{\`o}}, {Siskind}, {Smith},
  {Smith}, {Spandre}, {Spinelli}, {Strickman}, {Strong}, {Suson}, {Tajima},
  {Takahashi}, {Takahashi}, {Tanaka}, {Thayer}, {Thayer}, {Thompson},
  {Tibaldo}, {Torres}, {Tosti}, {Tramacere}, {Uchiyama}, {Usher}, {Van Etten},
  {Vasileiou}, {Venter}, {Vilchez}, {Vitale}, {Waite}, {Wang}, {Winer}, {Wood},
  {Ylinen}, \& {Ziegler}}]{ic443ferm}
---. 2010{\natexlab{d}}, \apj, 712, 459

\bibitem[{{Abdo} {et~al.}(2011){Abdo}, {Ackermann}, {Ajello}, {Allafort},
  {Baldini}, {Ballet}, {Barbiellini}, {Baring}, {Bastieri}, {Bellazzini},
  {Berenji}, {Blandford}, {Bloom}, {Bonamente}, {Borgland}, {Bouvier},
  {Brandt}, {Bregeon}, {Brigida}, {Bruel}, {Buehler}, {Buson}, {Caliandro},
  {Cameron}, {Caraveo}, {Casandjian}, {Cecchi}, {Chaty}, {Chekhtman}, {Cheung},
  {Chiang}, {Cillis}, {Ciprini}, {Claus}, {Cohen-Tanugi}, {Conrad}, {Corbel},
  {Cutini}, {de Angelis}, {de Palma}, {Dermer}, {Digel}, {Silva}, {Drell},
  {Drlica-Wagner}, {Dubois}, {Dumora}, {Favuzzi}, {Ferrara}, {Fortin},
  {Frailis}, {Fukazawa}, {Fukui}, {Funk}, {Fusco}, {Gargano}, {Gasparrini},
  {Gehrels}, {Germani}, {Giglietto}, {Giordano}, {Giroletti}, {Glanzman},
  {Godfrey}, {Grenier}, {Grondin}, {Guiriec}, {Hadasch}, {Hanabata}, {Harding},
  {Hayashida}, {Hayashi}, {Hays}, {Horan}, {Jackson}, {J{\'o}hannesson},
  {Johnson}, {Kamae}, {Katagiri}, {Kataoka}, {Kerr}, {Kn{\"o}dlseder}, {Kuss},
  {Lande}, {Latronico}, {Lee}, {Lemoine-Goumard}, {Longo}, {Loparco},
  {Lovellette}, {Lubrano}, {Madejski}, {Makeev}, {Mazziotta}, {McEnery},
  {Michelson}, {Mignani}, {Mitthumsiri}, {Mizuno}, {Moiseev}, {Monte},
  {Monzani}, {Morselli}, {Moskalenko}, {Murgia}, {Naumann-Godo}, {Nolan},
  {Norris}, {Nuss}, {Ohsugi}, {Okumura}, {Orlando}, {Ormes}, {Paneque},
  {Parent}, {Pelassa}, {Pesce-Rollins}, {Pierbattista}, {Piron}, {Pohl},
  {Porter}, {Rain{\`o}}, {Rando}, {Razzano}, {Reimer}, {Reposeur}, {Ritz},
  {Romani}, {Roth}, {Sadrozinski}, {Saz Parkinson}, {Sgr{\`o}}, {Smith},
  {Smith}, {Spandre}, {Spinelli}, {Strickman}, {Tajima}, {Takahashi},
  {Takahashi}, {Tanaka}, {Thayer}, {Thayer}, {Thompson}, {Tibaldo}, {Tibolla},
  {Torres}, {Tosti}, {Tramacere}, {Troja}, {Uchiyama}, {Vandenbroucke},
  {Vasileiou}, {Vianello}, {Vilchez}, {Vitale}, {Waite}, {Wang}, {Winer},
  {Wood}, {Yamamoto}, {Yamazaki}, {Yang}, \& {Ziegler}}]{rxferm}
---. 2011, \apj, 734, 28

\bibitem[{{Abdo} {et~al.}(2013){Abdo}, {Ajello}, {Allafort}, {Baldini},
  {Ballet}, {Barbiellini}, {Baring}, {Bastieri}, \& et~al.}]{2PC}
---. 2013, \apjs, 208, 17

\bibitem[{{Acciari} {et~al.}(2011){Acciari}, {Aliu}, {Arlen}, {Aune},
  {Beilicke}, {Benbow}, {Bradbury}, {Buckley}, {Bugaev}, {Byrum}, {Cannon},
  {Cesarini}, {Ciupik}, {Collins-Hughes}, {Cui}, {Dickherber}, {Duke},
  {Errando}, {Finley}, {Finnegan}, {Fortson}, {Furniss}, {Galante}, {Gall},
  {Gillanders}, {Godambe}, {Griffin}, {Grube}, {Guenette}, {Gyuk}, {Hanna},
  {Holder}, {Hughes}, {Hui}, {Humensky}, {Kaaret}, {Karlsson}, {Kertzman},
  {Kieda}, {Krawczynski}, {Krennrich}, {Lang}, {LeBohec}, {Madhavan}, {Maier},
  {Majumdar}, {McArthur}, {McCann}, {Moriarty}, {Mukherjee}, {Ong}, {Orr},
  {Otte}, {Pandel}, {Park}, {Perkins}, {Pohl}, {Quinn}, {Ragan}, {Reyes},
  {Reynolds}, {Roache}, {Rose}, {Saxon}, {Schroedter}, {Sembroski}, {Senturk},
  {Slane}, {Smith}, {Te{\v s}i{\'c}}, {Theiling}, {Thibadeau}, {Tsurusaki},
  {Varlotta}, {Vassiliev}, {Vincent}, {Vivier}, {Wakely}, {Ward}, {Weekes},
  {Weinstein}, {Weisgarber}, {Williams}, {Wood}, \& {Zitzer}}]{tyver}
{Acciari}, V.~A., {et~al.} 2011, \apjl, 730, L20

\bibitem[{{Ackermann} {et~al.}(2011){Ackermann}, {Ajello}, {Allafort},
  {Baldini}, {Ballet}, {Barbiellini}, {Bastieri}, {Belfiore}, {Bellazzini},
  {Berenji}, {Blandford}, {Bloom}, {Bonamente}, {Borgland}, {Bottacini},
  {Brigida}, {Bruel}, {Buehler}, {Buson}, {Caliandro}, {Cameron}, {Caraveo},
  {Casandjian}, {Cecchi}, {Chekhtman}, {Cheung}, {Chiang}, {Ciprini}, {Claus},
  {Cohen-Tanugi}, {de Angelis}, {de Palma}, {Dermer}, {do Couto e Silva},
  {Drell}, {Dumora}, {Favuzzi}, {Fegan}, {Focke}, {Fortin}, {Fukazawa},
  {Fusco}, {Gargano}, {Germani}, {Giglietto}, {Giordano}, {Giroletti},
  {Glanzman}, {Godfrey}, {Grenier}, {Guillemot}, { Guiriec}, {Hadasch},
  {Hanabata}, {Harding}, {Hayashida}, {Hayashi}, {Hays}, {J{\'o}hannesson},
  {Johnson}, {Kamae}, {Katagiri}, {Kataoka}, {Kerr}, {Kn{\"o}dlseder}, {Kuss},
  {Lande}, {Latronico}, {Lee}, {Longo}, {Loparco}, {Lott}, {Lovellette},
  {Lubrano}, {Martin}, {Mazziotta}, {McEnery}, {Mehault}, {Michelson},
  {Mitthumsiri}, {Mizuno}, {Monte}, {Monzani}, {Morselli}, {Moskalenko},
  {Murgia}, {Naumann-Godo}, {Nolan}, {Norris}, {Nuss}, {Ohsugi}, {Okumura},
  {Orlando}, {Ormes}, {Ozaki}, {Paneque}, {Parent}, {Pesce-Rollins},
  {Pierbattista}, {Piron}, {Pohl}, {Prokhorov}, {Rain{\`o}}, {Rando},
  {Razzano}, { Reposeur}, {Ritz}, {Parkinson}, {Sgr{\`o}}, {Siskind}, {Smith},
  {Spinelli}, {Strong}, {Takahashi}, {Tanaka}, {Thayer}, {Thayer}, {Thompson},
  {Tibaldo}, {Torres}, {Tosti}, {Tramacere}, {Troja}, {Uchiyama},
  {Vandenbroucke}, {Vasileiou}, {Vianello}, {Vitale}, {Waite}, {Wang}, {Winer},
  {Wood}, {Yang}, {Zimmer}, \& {Bontemps}}]{Cyg_cocoon}
{Ackermann}, M., {et~al.} 2011, Science, 334, 1103

\bibitem[{{Ackermann} {et~al.}(2012{\natexlab{a}}){Ackermann}, {Ajello},
  {Atwood}, {Baldini}, {Ballet}, {Barbiellini}, {Bastieri}, {Bechtol},
  {Bellazzini}, {Berenji}, {Blandford}, {Bloom}, {Bonamente}, {Borgland},
  {Brandt}, {Bregeon}, {Brigida}, {Bruel}, {Buehler}, {Buson}, {Caliandro},
  {Cameron}, {Caraveo}, {Cavazzuti}, {Cecchi}, {Charles}, {Chekhtman},
  {Chiang}, {Ciprini}, {Claus}, {Cohen-Tanugi}, {Conrad}, {Cutini}, {de
  Angelis}, {de Palma}, {Dermer}, {Digel}, {Silva}, {Drell}, {Drlica-Wagner},
  {Falletti}, {Favuzzi}, {Fegan}, {Ferrara}, {Focke}, {Fortin}, {Fukazawa},
  {Funk}, {Fusco}, {Gaggero}, {Gargano}, {Germani}, { Giglietto}, {Giordano},
  {Giroletti}, {Glanzman}, {Godfrey}, {Grove}, {Guiriec}, {Gustafsson},
  {Hadasch}, {Hanabata}, {Harding}, {Hayashida}, {Hays}, {Horan}, {Hou},
  {Hughes}, {J{\'o}hannesson}, {Johnson}, {Johnson}, {Kamae}, {Katagiri},
  {Kataoka}, {Kn{\"o}dlseder}, {Kuss}, {Lande}, {Latronico}, {Lee},
  {Lemoine-Goumard}, {Longo}, {Loparco}, {Lott}, {Lovellette}, {Lubrano},
  {Mazziotta}, {McEnery}, {Michelson}, {Mitthumsiri}, {Mizuno}, {Monte},
  {Monzani}, {Morselli}, {Moskalenko}, {Murgia}, {Naumann-Godo}, {Norris},
  {Nuss}, {Ohsugi}, {Okumura}, {Omodei}, {Orlando}, {Ormes}, {Paneque},
  {Panetta}, {Parent}, {Pesce-Rollins}, {Pierbattista}, {Piron}, {Pivato},
  {Porter}, {Rain{\`o}}, {Rando}, {Razzano}, {Razzaque}, {Reimer}, {Reimer},
  {Sadrozinski}, {Sgr{\`o}}, {Siskind}, {Spandre}, {Spinelli}, {Strong},
  {Suson}, {Takahashi}, {Tanaka}, {Thayer}, {Thayer}, {Thompson}, {Tibaldo},
  {Tinivella}, {Torres}, {Tosti}, {Troja}, {Usher}, {Vandenbroucke},
  {Vasileiou}, {Vianello}, {Vitale}, {Waite}, {Wang}, {Winer}, {Wood}, {Wood},
  {Yang}, {Ziegler}, \& {Zimmer}}]{diffpapII}
---. 2012{\natexlab{a}}, \apj, 750, 3

\bibitem[{{Ackermann} {et~al.}(2012{\natexlab{b}}){Ackermann}, {Ajello},
  {Allafort}, {Atwood}, {Axelsson}, {Baldini}, {Barbiellini}, {Bastieri},
  {Bechtol}, {Bellazzini}, {Berenji}, {Bloom}, {Bonamente}, {Borgland},
  {Bouvier}, {Bregeon}, {Brez}, {Brigida}, {Bruel}, {Buehler}, {Buson},
  {Caliandro}, {Cameron}, {Caraveo}, {Casandjian}, {Cecchi}, {Charles},
  {Chekhtman}, {Chiang}, {Ciprini}, {Claus}, {Cohen-Tanugi}, {Cutini}, {de
  Palma}, {Dermer}, {Digel}, {Do Couto E Silva}, {Drell}, {Drlica-Wagner},
  {Dubois}, {Enoto}, {Falletti}, {Favuzzi}, {Fegan}, {Focke}, {Fortin},
  {Fukazawa}, {Funk}, {Fusco}, {Gargano}, {Gehrels}, {Germani}, {Giglietto},
  {Giordano}, {Giroletti}, {Glanzman}, {Godfrey}, {Grenier}, {Grove},
  {Guiriec}, {Hadasch}, {Hayashida}, {Hays}, {Hughes}, {J{\'o}hannesson},
  {Johnson}, {Johnson}, {Kamae}, {Katagiri}, {Kataoka}, {Kn{\"o}dlseder},
  {Kuss}, {Lande}, {Latronico}, {Lee}, {Longo}, {Loparco}, {Lovellette},
  {Lubrano}, {Madejski}, {Mazziotta}, {McEnery}, {Michelson}, {Mizuno},
  {Moiseev}, {Monte}, {Monzani}, {Morselli}, {Moskalenko}, {Murgia},
  {Nakamori}, {Naumann-Godo}, {Nolan}, {Norris}, {Nuss}, {Ohsugi}, {Okumura},
  {Omodei}, {Orlando}, {Ormes}, {Ozaki}, {Paneque}, {Panetta}, {Parent},
  {Pesce-Rollins}, {Pierbattista}, {Piron}, {Rain{\`o}}, {Rando}, {Razzano},
  {Reimer}, {Reimer}, {Reposeur}, {Ritz}, {Rochester}, {Sgr{\`o}}, {Siskind},
  {Smith}, {Spandre}, {Spinelli}, {Suson}, {Takahashi}, {Tanaka}, {Thayer},
  {Thayer}, {Thompson}, {Tibaldo}, {Tosti}, {Troja}, {Usher}, {Vandenbroucke},
  {Vasileiou}, {Vianello}, {Vilchez}, {Vitale}, {Waite}, {Wang}, {Winer},
  {Wood}, {Yang}, \& {Zimmer}}]{bal12}
---. 2012{\natexlab{b}}, Astroparticle Physics, 35, 346

\bibitem[{{Ackermann} {et~al.}(2012{\natexlab{c}}){Ackermann}, {Ajello},
  {Allafort}, {Baldini}, {Ballet}, {Barbiellini}, {Bastieri}, {Belfiore},
  {Bellazzini}, {Berenji}, {Blandford}, {Bloom}, {Bonamente}, {Borgland},
  {Bottacini}, {Bregeon}, {Brigida}, {Bruel}, {Buehler}, {Buson}, {Caliandro},
  {Cameron}, {Caraveo}, {Casandjian}, {Cecchi}, {Chekhtman}, {Ciprini},
  {Claus}, {Cohen-Tanugi}, {de Angelis}, {de Palma}, {Dermer}, {Silva},
  {Drell}, {Dumora}, {Favuzzi}, {Fegan}, {Focke}, {Fortin}, {Fukazawa},
  {Fusco}, {Gargano}, {Germani}, {Giglietto}, {Giordano}, {Giroletti},
  {Glanzman}, {Godfrey}, {Grenier}, {Guillemot}, {Guiriec}, { Hadasch},
  {Hanabata}, {Harding}, {Hayashida}, {Hayashi}, {Hays}, {J{\'o}hannesson},
  {Johnson}, {Kamae}, {Katagiri}, {Kataoka}, {Kerr}, {Kn{\"o}dlseder}, {Kuss},
  {Lande}, {Latronico}, {Lee}, {Longo}, {Loparco}, {Lott}, {Lovellette},
  {Lubrano}, {Martin}, {Mazziotta}, {McEnery}, {Mehault}, {Michelson},
  {Mitthumsiri}, {Mizuno}, {Monte}, {Monzani}, {Morselli}, {Moskalenko},
  {Murgia}, {Naumann-Godo}, {Nolan}, {Norris}, {Nuss}, {Ohsugi}, {Okumura},
  {Omodei}, {Orlando}, {Ormes}, {Ozaki}, {Paneque}, {Parent}, {Pesce-Rollins},
  {Pierbattista}, {Piron}, {Porter}, {Rain{\`o}}, {Rando}, {Razzano}, {Reimer},
  { Reposeur}, {Ritz}, {Saz Parkinson}, {Sgr{\`o}}, {Siskind}, {Smith},
  {Spinelli}, {Strong}, {Takahashi}, {Tanaka}, {Thayer}, {Thayer}, {Thompson},
  {Tibaldo}, {Torres}, {Tosti}, {Tramacere}, {Troja}, {Uchiyama},
  {Vandenbroucke}, {Vasileiou}, {Vianello}, {Vitale}, {Waite}, {Wang}, {Winer},
  {Wood}, {Yang}, {Zimmer}, \& {Bontemps}}]{LATCygISM2012}
---. 2012{\natexlab{c}}, \aap, 538, A71

\bibitem[{{Ackermann} {et~al.}(2013{\natexlab{a}}){Ackermann}, {Ajello},
  {Allafort}, {Baldini}, {Ballet}, {Barbiellini}, {Baring}, {Bastieri},
  {Bechtol}, {Bellazzini}, {Blandford}, {Bloom}, {Bonamente}, {Borgland},
  {Bottacini}, {Brandt}, {Bregeon}, {Brigida}, {Bruel}, {Buehler}, {Busetto},
  {Buson}, {Caliandro}, {Cameron}, {Caraveo}, {Casandjian}, {Cecchi}, {{\c
  C}elik}, {Charles}, {Chaty}, {Chaves}, {Chekhtman}, {Cheung}, {Chiang},
  {Chiaro}, {Cillis}, {Ciprini}, {Claus}, {Cohen-Tanugi}, {Cominsky}, {Conrad},
  {Corbel}, {Cutini}, {D'Ammando}, {de Angelis}, {de Palma}, {Dermer}, {do
  Couto e Silva}, {Drell}, {Drlica-Wagner}, {Falletti}, {Favuzzi}, {Ferrara},
  {Franckowiak}, {Fukazawa}, {Funk}, {Fusco}, {Gargano}, {Germani},
  {Giglietto}, {Giommi}, {Giordano}, {Giroletti}, {Glanzman}, {Godfrey},
  {Grenier}, {Grondin}, {Grove}, {Guiriec}, {Hadasch}, {Hanabata}, {Harding},
  {Hayashida}, {Hayashi}, {Hays}, {Hewitt}, {Hill}, {Hughes}, {Jackson},
  {Jogler}, {J{\'o}hannesson}, {Johnson}, {Kamae}, {Kataoka}, {Katsuta},
  {Kn{\"o}dlseder}, {Kuss}, {Lande}, {Larsson}, {Latronico}, {Lemoine-Goumard},
  {Longo}, {Loparco}, {Lovellette}, {Lubrano}, {Madejski}, {Massaro}, {Mayer},
  {Mazziotta}, {McEnery}, {Mehault}, {Michelson}, {Mignani}, {Mitthumsiri},
  {Mizuno}, {Moiseev}, {Monzani}, {Morselli}, {Moskalenko}, {Murgia},
  {Nakamori}, {Nemmen}, {Nuss}, {Ohno}, {Ohsugi}, {Omodei}, {Orienti},
  {Orlando}, {Ormes}, {Paneque}, {Perkins}, {Pesce-Rollins}, {Piron}, {Pivato},
  {Rain{\`o}}, {Rando}, {Razzano}, {Razzaque}, {Reimer}, {Reimer}, {Ritz},
  {Romoli}, {S{\'a}nchez-Conde}, {Schulz}, {Sgr{\`o}}, {Simeon}, {Siskind},
  {Smith}, {Spandre}, {Spinelli}, {Stecker}, {Strong}, {Suson}, {Tajima},
  {Takahashi}, {Takahashi}, {Tanaka}, {Thayer}, {Thayer}, {Thompson},
  {Thorsett}, {Tibaldo}, {Tibolla}, {Tinivella}, {Troja}, {Uchiyama}, {Usher},
  {Vandenbroucke}, {Vasileiou}, {Vianello}, {Vitale}, {Waite}, {Werner},
  {Winer}, {Wood}, {Wood}, {Yamazaki}, {Yang}, \& {Zimmer}}]{fermipidec}
---. 2013{\natexlab{a}}, Science, 339, 807

\bibitem[{{Ackermann} {et~al.}(2013{\natexlab{b}}){Ackermann}, {Ajello},
  {Allafort}, {Baldini}, {Ballet}, {Barbiellini}, {Baring}, {Bastieri},
  {Bechtol}, {Bellazzini}, {Blandford}, {Bloom}, {Bonamente}, {Borgland},
  {Bottacini}, {Brandt}, {Bregeon}, {Brigida}, {Bruel}, {Buehler}, {Busetto},
  {Buson}, {Caliandro}, {Cameron}, {Caraveo}, {Casandjian}, {Cecchi}, {{\c
  C}elik}, {Charles}, {Chaty}, {Chaves}, {Chekhtman}, {Cheung}, {Chiang},
  {Chiaro}, {Cillis}, {Ciprini}, {Claus}, {Cohen-Tanugi}, {Cominsky}, {Conrad},
  {Corbel}, {Cutini}, {D'Ammando}, {de Angelis}, {de Palma}, {Dermer}, {do
  Couto e Silva}, {Drell}, {Drlica-Wagner}, {Falletti}, {Favuzzi}, {Ferrara},
  {Franckowiak}, {Fukazawa}, {Funk}, {Fusco}, {Gargano}, {Germani},
  {Giglietto}, {Giommi}, {Giordano}, {Giroletti}, {Glanzman}, {Godfrey},
  {Grenier}, {Grondin}, {Grove}, {Guiriec}, {Hadasch}, {Hanabata}, {Harding},
  {Hayashida}, {Hayashi}, {Hays}, {Hewitt}, {Hill}, {Hughes}, {Jackson},
  {Jogler}, {J{\'o}hannesson}, {Johnson}, {Kamae}, {Kataoka}, {Katsuta},
  {Kn{\"o}dlseder}, {Kuss}, {Lande}, {Larsson}, {Latronico}, {Lemoine-Goumard},
  {Longo}, {Loparco}, {Lovellette}, {Lubrano}, {Madejski}, {Massaro}, {Mayer},
  {Mazziotta}, {McEnery}, {Mehault}, {Michelson}, {Mignani}, {Mitthumsiri},
  {Mizuno}, {Moiseev}, {Monzani}, {Morselli}, {Moskalenko}, {Murgia},
  {Nakamori}, {Nemmen}, {Nuss}, {Ohno}, {Ohsugi}, {Omodei}, {Orienti},
  {Orlando}, {Ormes}, {Paneque}, {Perkins}, {Pesce-Rollins}, {Piron}, {Pivato},
  {Rain{\`o}}, {Rando}, {Razzano}, {Razzaque}, {Reimer}, {Reimer}, {Ritz},
  {Romoli}, {S{\'a}nchez-Conde}, {Schulz}, {Sgr{\`o}}, {Simeon}, {Siskind},
  {Smith}, {Spandre}, {Spinelli}, {Stecker}, {Strong}, {Suson}, {Tajima},
  {Takahashi}, {Takahashi}, {Tanaka}, {Thayer}, {Thayer}, {Thompson},
  {Thorsett}, {Tibaldo}, {Tibolla}, {Tinivella}, {Troja}, {Uchiyama}, {Usher},
  {Vandenbroucke}, {Vasileiou}, {Vianello}, {Vitale}, {Waite}, {Werner},
  {Winer}, {Wood}, {Wood}, {Yamazaki}, {Yang}, \& {Zimmer}}]{pionbump}
---. 2013{\natexlab{b}}, Science, 339, 807

\bibitem[{{Aharonian} {et~al.}(2007){Aharonian}, {Akhperjanian}, {Bazer-Bachi},
  {Beilicke}, {Benbow}, {Berge}, {Bernl{\"o}hr}, {Boisson}, {Bolz}, {Borrel},
  {Braun}, {Brion}, {Brown}, {B{\"u}hler}, {B{\"u}sching}, {Carrigan},
  {Chadwick}, {Chounet}, {Coignet}, {Cornils}, {Costamante}, {Degrange},
  {Dickinson}, {Djannati-Ata{\"i}}, {O'C.~Drury}, {Dubus}, {Egberts},
  {Emmanoulopoulos}, {Espigat}, {Feinstein}, {Ferrero}, {Fiasson}, {Fontaine},
  {Funk}, {Funk}, {F{\"u}{\ss}ling}, {Gallant}, {Giebels}, {Glicenstein},
  {Gl{\"u}ck}, {Goret}, {Hadjichristidis}, {Hauser}, {Hauser}, {Heinzelmann},
  {Henri}, {Hermann}, {Hinton}, {Hoffmann}, {Hofmann}, {Holleran}, {Hoppe},
  {Horns}, {Jacholkowska}, {de Jager}, {Kendziorra}, {Kerschhaggl},
  {Kh{\'e}lifi}, {Komin}, {Konopelko}, {Kosack}, {Lamanna}, {Latham}, {Le
  Gallou}, {Lemi{\`e}re}, {Lemoine-Goumard}, {Lohse}, {Martin},
  {Martineau-Huynh}, {Marcowith}, {Masterson}, {Maurin}, {McComb}, {Moulin},
  {de Naurois}, {Nedbal}, {Nolan}, {Noutsos}, {Olive}, {Orford}, {Osborne},
  {Panter}, {Pelletier}, {Pita}, {P{\"u}hlhofer}, {Punch}, {Ranchon},
  {Raubenheimer}, {Raue}, {Rayner}, {Reimer}, {Reimer}, {Ripken}, {Rob},
  {Rolland}, {Rosier-Lees}, {Rowell}, {Sahakian}, {Santangelo}, {Saug{\'e}},
  {Schlenker}, {Schlickeiser}, {Schr{\"o}der}, { Schwanke}, {Schwarzburg},
  {Schwemmer}, {Shalchi}, {Sol}, {Spangler}, {Spanier}, {Steenkamp},
  {Stegmann}, {Superina}, {Tam}, {Tavernet}, {Terrier}, {Tluczykont}, {van
  Eldik}, {Vasileiadis}, {Venter}, {Vialle}, {Vincent}, {V{\"o}lk}, {Wagner},
  \& {Ward}}]{rxhess}
{Aharonian}, F., {et~al.} 2007, \aap, 464, 235

\bibitem[{{Aharonian} \& {Atoyan}(1996)}]{aharonian96}
{Aharonian}, F.~A., \& {Atoyan}, A.~M. 1996, \aap, 309, 917

\bibitem[{{Albert} {et~al.}(2007){Albert}, {Aliu}, {Anderhub}, {Antoranz},
  {Armada}, {Baixeras}, {Barrio}, {Bartko}, {Bastieri}, {Becker}, {Bednarek},
  {Berger}, {Bigongiari}, {Biland}, {Bock}, {Bordas}, {Bosch-Ramon}, {Bretz},
  {Britvitch}, {Camara}, {Carmona}, {Chilingarian}, {Coarasa}, {Commichau},
  {Contreras}, {Cortina}, {Costado}, {Curtef}, {Danielyan}, {Dazzi}, {De
  Angelis}, {Delgado}, {de los Reyes}, {De Lotto}, {Domingo-Santamar{\'{\i}}a},
  {Dorner}, {Doro}, {Errando}, {Fagiolini}, {Ferenc}, {Fern{\'a}ndez}, {Firpo},
  {Flix}, {Fonseca}, {Font}, {Fuchs}, {Galante}, {Garc{\'{\i}}a-L{\'o}pez},
  {Garczarczyk}, {Gaug}, {Giller}, {Goebel}, {Hakobyan}, {Hayashida},
  {Hengstebeck}, {Herrero}, {H{\"o}hne}, {Hose}, {Hsu}, {Jacon}, {Jogler},
  {Kosyra}, {Kranich}, {Kritzer}, {Laille}, {Lindfors}, {Lombardi}, {Longo},
  {L{\'o}pez}, {L{\'o}pez}, {Lorenz}, {Majumdar}, {Maneva}, {Mannheim},
  {Mansutti}, {Mariotti}, {Mart{\'{\i}}nez}, {Mazin}, {Merck}, {Meucci},
  {Meyer}, {Miranda}, {Mirzoyan}, {Mizobuchi}, {Moralejo}, {Nieto}, {Nilsson},
  {Ninkovic}, {O{\~n}a-Wilhelmi}, {Otte}, {Oya}, {Paneque}, {Panniello},
  {Paoletti}, {Paredes}, {Pasanen}, {Pascoli}, {Pauss}, {Pegna}, {Persic},
  {Peruzzo}, {Piccioli}, {Prandini}, {Puchades}, {Raymers}, {Rhode},
  {Rib{\'o}}, {Rico}, {Rissi}, {Robert}, {R{\"u}gamer}, {Saggion}, {Saito},
  {S{\'a}nchez}, {Sartori}, {Scalzotto}, {Scapin}, {Schmitt}, {Schweizer},
  {Shayduk}, {Shinozaki}, {Shore}, {Sidro}, {Sillanp{\"a}{\"a}}, {Sobczynska},
  {Stamerra}, {Stark}, {Takalo}, {Temnikov}, {Tescaro}, {Teshima}, {Torres},
  {Turini}, {Vankov}, {Vitale}, {Wagner}, {Wibig}, {Wittek}, {Zandanel},
  {Zanin}, \& {Zapatero}}]{ic443mag}
{Albert}, J., {et~al.} 2007, \apjl, 664, L87

\bibitem[{{Atwood} {et~al.}(2009){Atwood}, {Abdo}, {Ackermann}, {Althouse},
  {Anderson}, {Axelsson}, {Baldini}, {Ballet}, {Band}, {Barbiellini}, \&
  et~al.}]{Atw09}
{Atwood}, W.~B., {et~al.} 2009, \apj, 697, 1071

\bibitem[{{Byun} {et~al.}(2006){Byun}, {Koo}, {Tatematsu}, \& {Sunada}}]{Byu06}
{Byun}, D.-Y., {Koo}, B.-C., {Tatematsu}, K., \& {Sunada}, K. 2006, \apj, 637,
  283

\bibitem[{{Case} \& {Bhattacharya}(1998)}]{case1998}
{Case}, G.~L., \& {Bhattacharya}, D. 1998, \apj, 504, 761

\bibitem[{{Castro} \& {Slane}(2010)}]{3c391fermi}
{Castro}, D., \& {Slane}, P. 2010, \apj, 717, 372

\bibitem[{{Dame} {et~al.}(2001){Dame}, {Hartmann}, \& {Thaddeus}}]{dame01}
{Dame}, T.~M., {Hartmann}, D., \& {Thaddeus}, P. 2001, \apj, 547, 792

\bibitem[{{Dame} \& {Thaddeus}(2011)}]{dame11}
{Dame}, T.~M., \& {Thaddeus}, P. 2011, \apjl, 734, L24

\bibitem[{{de Palma} {et~al.}(2013){de Palma}, {Brandt}, {Johannesson},
  {Tibaldo}, \& {for the Fermi LAT collaboration}}]{depalma13}
{de Palma}, F., {Brandt}, T.~J., {Johannesson}, G., {Tibaldo}, L., \& {for the
  Fermi LAT collaboration}. 2013, ArXiv e-prints

\bibitem[{{Drury}(2012)}]{drury12}
{Drury}, L.~O.~. 2012, Astroparticle Physics, 39, 52

\bibitem[{{Flower} \& {Pineau des For{\^e}ts}(1999)}]{flo99}
{Flower}, D.~R., \& {Pineau des For{\^e}ts}, G. 1999, \mnras, 308, 271

\bibitem[{{Gabici} {et~al.}(2009){Gabici}, {Aharonian}, \&
  {Casanova}}]{gabici09}
{Gabici}, S., {Aharonian}, F.~A., \& {Casanova}, S. 2009, \mnras, 396, 1629

\bibitem[{{Gaisser}(1990)}]{gaisser90}
{Gaisser}, T.~K. 1990, {Cosmic rays and particle physics}

\bibitem[{{Gao} {et~al.}(2011){Gao}, {Han}, {Reich}, {Reich}, {Sun}, \&
  {Xiao}}]{gao11}
{Gao}, X.~Y., {Han}, J.~L., {Reich}, W., {Reich}, P., {Sun}, X.~H., \& {Xiao},
  L. 2011, \aap, 529, A159

\bibitem[{{Giordano} {et~al.}(2012){Giordano}, {Naumann-Godo}, {Ballet},
  {Bechtol}, {Funk}, {Lande}, {Mazziotta}, {Rain{\`o}}, {Tanaka}, {Tibolla}, \&
  {Uchiyama}}]{tyferm}
{Giordano}, F., {et~al.} 2012, \apjl, 744, L2

\bibitem[{{Giuliani} {et~al.}(2011){Giuliani}, {Cardillo}, {Tavani}, {Fukui},
  {Yoshiike}, {Torii}, {Dubner}, {Castelletti}, {Barbiellini}, {Bulgarelli},
  {Caraveo}, {Costa}, {Cattaneo}, {Chen}, {Contessi}, {Del Monte},
  {Donnarumma}, {Evangelista}, {Feroci}, {Gianotti}, {Lazzarotto}, {Lucarelli},
  {Longo}, {Marisaldi}, {Mereghetti}, {Pacciani}, {Pellizzoni}, {Piano},
  {Picozza}, {Pittori}, {Pucella}, {Rapisarda}, {Rappoldi}, {Sabatini},
  {Soffitta}, {Striani}, {Trifoglio}, {Trois}, {Vercellone}, {Verrecchia},
  {Vittorini}, {Colafrancesco}, {Giommi}, \& {Bignami}}]{w44ag}
{Giuliani}, A., {et~al.} 2011, \apjl, 742, L30

\bibitem[{{Green}(2009)}]{green}
{Green}, D.~A. 2009, Bulletin of the Astronomical Society of India, 37, 45

\bibitem[{{Grenier} {et~al.}(2005){Grenier}, {Casandjian}, \&
  {Terrier}}]{grenier2005}
{Grenier}, I.~A., {Casandjian}, J.-M., \& {Terrier}, R. 2005, Science, 307,
  1292

\bibitem[{{Hewitt} {et~al.}(2012){Hewitt}, {Grondin}, {Goumard-Lemoine},
  {Reposuer}, {Ballet}, \& {others}}]{hewitt12}
{Hewitt}, J.~W., {Grondin}, M.-H., {Goumard-Lemoine}, M., {Reposuer}, T.,
  {Ballet}, J., \& {others}. 2012, \apj, in print

\bibitem[{{Hewitt} {et~al.}(2009){Hewitt}, {Yusef-Zadeh}, \& {Wardle}}]{hew09}
{Hewitt}, J.~W., {Yusef-Zadeh}, F., \& {Wardle}, M. 2009, \apjl, 706, L270

\bibitem[{{Hill}(1974)}]{hill74}
{Hill}, L.~E. 1974, \mnras, 169, 59

\bibitem[{{Houck} \& {Allen}(2006)}]{houck2006}
{Houck}, J.~C., \& {Allen}, G.~E. 2006, \apjs, 167, 26

\bibitem[{{Houck} \& {Denicola}(2000)}]{isis}
{Houck}, J.~C., \& {Denicola}, L.~A. 2000, in Astronomical Society of the
  Pacific Conference Series, Vol. 216, Astronomical Data Analysis Software and
  Systems IX, ed. N.~{Manset}, C.~{Veillet}, \& D.~{Crabtree}, 591

\bibitem[{{Janssen} {et~al.}(2009){Janssen}, {Stappers}, {Braun}, {van
  Straten}, {Edwards}, {Rubio-Herrera}, {van Leeuwen}, \& {Weltevrede}}]{pul09}
{Janssen}, G.~H., {Stappers}, B.~W., {Braun}, R., {van Straten}, W., {Edwards},
  R.~T., {Rubio-Herrera}, E., {van Leeuwen}, J., \& {Weltevrede}, P. 2009,
  \aap, 498, 223

\bibitem[{{Jarosik} {et~al.}(2011){Jarosik}, {Bennett}, {Dunkley}, {Gold},
  {Greason}, {Halpern}, {Hill}, {Hinshaw}, {Kogut}, {Komatsu}, {Larson},
  {Limon}, {Meyer}, {Nolta}, {Odegard}, {Page}, {Smith}, {Spergel}, {Tucker},
  {Weiland}, {Wollack}, \& {Wright}}]{Jarosik2011}
{Jarosik}, N., {et~al.} 2011, \apjs, 192, 14

\bibitem[{{Karlsson} \& {Kamae}(2008)}]{Karlsson2008}
{Karlsson}, N., \& {Kamae}, T. 2008, \apj, 674, 278

\bibitem[{Katagiri {et~al.}(2011)Katagiri, Tibaldo, Ballet, Giordano, Grenier,
  Porter, Roth, Tibolla, Uchiyama, \& Yamazaki}]{cyg11}
Katagiri, H., {et~al.} 2011, The Astrophysical Journal, 741, 44

\bibitem[{{Katsuta} {et~al.}(2012){Katsuta}, {Uchiyama}, {Tanaka}, {Tajima},
  {Bechtol}, {Funk}, {Lande}, {Ballet}, {Hanabata}, {Lemoine-Goumard}, \&
  {Takahashi}}]{katsuta12}
{Katsuta}, J., {et~al.} 2012, \apj, 752, 135

\bibitem[{{Knoedlseder} {et~al.}(1996){Knoedlseder}, {Oberlack}, {Diehl},
  {Chen}, \& {Gehrels}}]{kno96}
{Knoedlseder}, J., {Oberlack}, U., {Diehl}, R., {Chen}, W., \& {Gehrels}, N.
  1996, \aaps, 120, C339

\bibitem[{{Koo} \& {Heiles}(1991)}]{koo91}
{Koo}, B.-C., \& {Heiles}, C. 1991, \apj, 382, 204

\bibitem[{Koo {et~al.}(2001)Koo, Rho, Reach, Jung, \& Mangum}]{koo01}
Koo, B.-C., Rho, J., Reach, W.~T., Jung, J., \& Mangum, J.~G. 2001, The
  Astrophysical Journal, 552, 175

\bibitem[{{Kothes} {et~al.}(2006){Kothes}, {Fedotov}, {Foster}, \&
  {Uyan{\i}ker}}]{kothes06}
{Kothes}, R., {Fedotov}, K., {Foster}, T.~J., \& {Uyan{\i}ker}, B. 2006, \aap,
  457, 1081

\bibitem[{{Lande} {et~al.}(2012){Lande}, {Ackermann}, {Allafort}, {Ballet},
  {Bechtol}, {Burnett}, {Cohen-Tanugi}, {Drlica-Wagner}, {Funk}, {Giordano},
  {Grondin}, {Kerr}, \& {Lemoine-Goumard}}]{lan12}
{Lande}, J., {et~al.} 2012, \apj, 756, 5

\bibitem[{{Lazendic} \& {Slane}(2006)}]{laz06}
{Lazendic}, J.~S., \& {Slane}, P.~O. 2006, \apj, 647, 350

\bibitem[{{Leahy}(2006)}]{leahy06}
{Leahy}, D.~A. 2006, \apj, 647, 1125

\bibitem[{{Leahy} \& {Aschenbach}(1996)}]{lea96}
{Leahy}, D.~A., \& {Aschenbach}, B. 1996, \aap, 315, 260

\bibitem[{{Leahy} \& {Roger}(1998)}]{leahy98}
{Leahy}, D.~A., \& {Roger}, R.~S. 1998, \apj, 505, 784

\bibitem[{{Lorimer} {et~al.}(2006){Lorimer}, {Faulkner}, {Lyne}, {Manchester},
  {Kramer}, {McLaughlin}, {Hobbs}, {Possenti}, {Stairs}, {Camilo}, {Burgay},
  {D'Amico}, {Corongiu}, \& {Crawford}}]{lorimer2006}
{Lorimer}, D.~R., {et~al.} 2006, \mnras, 372, 777

\bibitem[{{Malkov} {et~al.}(2011){Malkov}, {Diamond}, \& {Sagdeev}}]{malkov11}
{Malkov}, M.~A., {Diamond}, P.~H., \& {Sagdeev}, R.~Z. 2011, Nature
  Communications, 2

\bibitem[{{Mattox} {et~al.}(1996){Mattox}, {Bertsch}, {Chiang}, {Dingus},
  {Digel}, {Esposito}, {Fierro}, {Hartman}, \& {others}}]{Mat96}
{Mattox}, J.~R., {et~al.} 1996, \apj, 461, 396

\bibitem[{Mavromatakis {et~al.}(2007)Mavromatakis, Xilouris, \&
  Boumis}]{mavromatakis07}
Mavromatakis, F., Xilouris, E.~M., \& Boumis, P. 2007, Astronomy and
  Astrophysics, 461, 991

\bibitem[{{Nolan} {et~al.}(2012){Nolan}, {Abdo}, {Ackermann}, {Ajello},
  {Allafort}, {Antolini}, {Atwood}, {Axelsson}, {Baldini}, {Ballet}, \&
  et~al.}]{2fgl}
{Nolan}, P.~L., {et~al.} 2012, \apjs, 199, 31

\bibitem[{{Protassov} {et~al.}(2002){Protassov}, {van Dyk}, {Connors},
  {Kashyap}, \& {Siemiginowska}}]{Protassov02}
{Protassov}, R., {van Dyk}, D.~A., {Connors}, A., {Kashyap}, V.~L., \&
  {Siemiginowska}, A. 2002, \apj, 571, 545

\bibitem[{{Reichardt} {et~al.}(2012){Reichardt}, {de O{\~n}a-Wilhelmi}, {Rico},
  \& {Yang}}]{reichardt12}
{Reichardt}, I., {de O{\~n}a-Wilhelmi}, E., {Rico}, J., \& {Yang}, R. 2012,
  \aap, 546, A21

\bibitem[{{Reynolds}(2008)}]{rey08}
{Reynolds}, S.~P. 2008, \araa, 46, 89

\bibitem[{{Reynolds}(2009)}]{rey09}
---. 2009, \apj, 703, 662

\bibitem[{{Rho} \& {Petre}(1998)}]{rho98}
{Rho}, J., \& {Petre}, R. 1998, \apjl, 503, L167

\bibitem[{{Shinn} {et~al.}(2009){Shinn}, {Koo}, {Burton}, {Lee}, \&
  {Moon}}]{shi09}
{Shinn}, J.-H., {Koo}, B.-C., {Burton}, M.~G., {Lee}, H.-G., \& {Moon}, D.-S.
  2009, \apj, 693, 1883

\bibitem[{{Smith} {et~al.}(2008){Smith}, {Guillemot}, {Camilo}, {Cognard},
  {Dumora}, {Espinoza}, {Freire}, {Gotthelf}, {Harding}, {Hobbs}, {Johnston},
  {Kaspi}, {Kramer}, {Livingstone}, {Lyne}, {Manchester}, {Marshall},
  {McLaughlin}, {Noutsos}, {Ransom}, {Roberts}, {Romani}, {Stappers},
  {Theureau}, {Thompson}, {Thorsett}, {Wang}, \& {Weltevrede}}]{smi08}
{Smith}, D.~A., {et~al.} 2008, \aap, 492, 923

\bibitem[{{Sturner} {et~al.}(1997){Sturner}, {Skibo}, {Dermer}, \&
  {Mattox}}]{sturner97}
{Sturner}, S.~J., {Skibo}, J.~G., {Dermer}, C.~D., \& {Mattox}, J.~R. 1997,
  \apj, 490, 619

\bibitem[{{Uchiyama} {et~al.}(2010){Uchiyama}, {Blandford}, {Funk}, {Tajima},
  \& {Tanaka}}]{uchiyama10}
{Uchiyama}, Y., {Blandford}, R.~D., {Funk}, S., {Tajima}, H., \& {Tanaka}, T.
  2010, \apjl, 723, L122

\bibitem[{{Voges} {et~al.}(1999){Voges}, {Aschenbach}, {Boller},
  {Br{\"a}uninger}, {Briel}, {Burkert}, {Dennerl}, {Englhauser}, {Gruber},
  {Haberl}, {Hartner}, {Hasinger}, {K{\"u}rster}, {Pfeffermann}, {Pietsch},
  {Predehl}, {Rosso}, {Schmitt}, {Tr{\"u}mper}, \& {Zimmermann}}]{rosat}
{Voges}, W., {et~al.} 1999, \aap, 349, 389

\bibitem[{{Wang} \& {Hirotani}(2011)}]{wan11}
{Wang}, R.-B., \& {Hirotani}, K. 2011, \apj, 736, 127

\bibitem[{{Webb} {et~al.}(1984){Webb}, {Drury}, \& {Biermann}}]{webb84}
{Webb}, G.~M., {Drury}, L.~O., \& {Biermann}, P. 1984, \aap, 137, 185

\bibitem[{{Weiland} {et~al.}(2011){Weiland}, {Odegard}, {Hill}, {Wollack},
  {Hinshaw}, {Greason}, {Jarosik}, {Page}, {Bennett}, {Dunkley}, {Gold},
  {Halpern}, {Kogut}, {Komatsu}, {Larson}, {Limon}, {Meyer}, {Nolta}, {Smith},
  {Spergel}, {Tucker}, \& {Wright}}]{weiland11}
{Weiland}, J.~L., {et~al.} 2011, \apjs, 192, 19

\bibitem[{{Yusef-Zadeh} {et~al.}(2003){Yusef-Zadeh}, {Wardle}, {Rho}, \&
  {Sakano}}]{zadeh03}
{Yusef-Zadeh}, F., {Wardle}, M., {Rho}, J., \& {Sakano}, M. 2003, \apj, 585,
  319

\end{thebibliography}

\end{document}